\documentclass[useAMS,usenatbib,fleqn]{mnras}

\usepackage{graphicx,color}
\usepackage{mathtools}	
\usepackage{amsmath}
\usepackage{amssymb}
\usepackage{bm}
\usepackage{subfig}             
\usepackage{ulem}
\usepackage{fixltx2e} 
\usepackage{pstricks} 

\def\aap{A\&A}
\def\apjl{ApJ}

\def\apjs{ApJS}

\newcommand{\Exp}[1]{{\rm e}^{#1}}

\newcommand{\del}{\partial}

\newcommand{\bfDel}{\bm{\nabla}}
\newcommand{\bmDel}{\bm{\nabla}}

\newcommand{\bfOmega}{\bm{\Omega}}

\newcommand{\alp}{\alpha}

\newcommand{\Emf}{\bm{\mathcal{E}}}

\newcommand{\Flux}{\bm{\mathcal{F}}}

\newcommand{\bfu}{\bm{u}}
\newcommand{\bfb}{\bm{b}}

\newcommand{\bfz}{\bm{z}}

\newcommand{\mean}[1]{{#1}}
\newcommand{\meanv}[1]{{\bm{#1}}}

\newcommand{\eq}{_\mathrm{eq}}						
\newcommand{\f}{_\mathrm{0}}					   	
\newcommand{\kin}{_\mathrm{k}}			   		
\newcommand{\magn}{_\mathrm{m}}			   		
\newcommand{\crit}{_\mathrm{c}}			   		
\newcommand{\sat}{_\mathrm{sat}}			   		
\newcommand{\pol}{_\mathrm{p}}			   		

\newcommand{\Adv}{\mathrm{a}}			   		
\newcommand{\Diff}{\mathrm{d}}			   		

\newcommand{\cro}{\times}

\newcommand{\mbr}{\mean{B}_r}
\newcommand{\mbp}{\mean{B}_\phi}
\newcommand{\mbz}{\mean{B}_z}
\newcommand{\mbi}{\mean{B}_i}
\newcommand{\mur}{\mean{U}_r}
\newcommand{\mup}{\mean{U}_\phi}
\newcommand{\muz}{\mean{U}_z}

\newcommand{\Btilde}{\widetilde{B}}
\newcommand{\mbrtilde}{\widetilde{B}_r}
\newcommand{\mbptilde}{\widetilde{B}_\phi}
\newcommand{\Dtilde}{\widetilde{D}}

\renewcommand{\theenumi}{\roman{enumi}}

\newcommand{\A}{_\mathrm{A}}

\newcommand{\Sz}{S_{z}}
\newcommand{\Coriolis}{\mathrm{Co}}
\newcommand{\Strouhal}{\mathrm{St}}

  \newcommand{\kms}{\,{\rm km\,s^{-1}}}

  \newcommand{\kpc}{\,{\rm kpc}}
  \newcommand{\pc}{\,{\rm pc}}
  
  \newcommand{\Myr}{\,{\rm Myr}}
  
  \newcommand{\Gyr}{\,{\rm Gyr}}

  \newcommand{\mkG}{\,\mu{\rm G}}

\title[An analytical dynamo solution for galaxies]{An analytical dynamo solution for large-scale magnetic fields of galaxies}
\author[L.\ Chamandy]{Luke Chamandy,$^{1,2}$\thanks{E-mail: luke@ast.uct.ac.za}\\
$^{1}$Astronomy Department, University of Cape Town, Rondebosch 7701, Republic of South Africa\\
$^{2}$Department of Physics, University of the Western Cape, Belleville 7535, Republic of South Africa
}

\begin{document}


\pagerange{\pageref{firstpage}--\pageref{lastpage}} \pubyear{2016}

\maketitle

\label{firstpage}
\begin{abstract}
We present an effectively global analytical asymptotic galactic dynamo solution 
for the regular magnetic field of an axisymmetric thin disc 
in the saturated state.
This solution is constructed by combining two well-known types of local galactic dynamo solution,
parameterized by the disc radius.
Namely, the critical (zero growth) solution obtained 
by treating the dynamo equation as a perturbed diffusion equation
is normalized using a non-linear solution that makes use of the `no-$z$' approximation 
and the dynamical $\alpha$-quenching non-linearity.
This overall solution is found to be reasonably accurate when compared 
with detailed numerical solutions.
It is thus potentially useful as a tool for predicting observational signatures 
of magnetic fields of galaxies.
In particular, such solutions could be painted onto galaxies in cosmological simulations
to enable the construction of synthetic polarized synchrotron and Faraday rotation measure (RM) datasets.
Further, we explore the properties of our numerical solutions,
and their dependence on certain parameter values.
We illustrate and assess the degree to which numerical solutions based on various levels of approximation,
common in the dynamo literature, agree with one another.
\end{abstract}

\section{Introduction}
\label{sec:intro}
With the advent of the Square Kilometre Array (SKA), 
magnetic fields will be regularly probed out to high redshift \citep{Taylor+15}.
It is necessary, therefore, to model theoretically such magnetic fields.
Much recent progress has been made in magnetohydrodynamical (MHD) cosmological simulations,
but the dynamic range allowed by such simulations is still too small to faithfully account 
for important (sub)galactic-scale processes including dynamo action.
Therefore, theoretical modelling of magnetic fields of galaxies, 
using a combination of analytical and numerical approaches, is crucial.
In particular, magnetic fields of magnitude $\sim10\mkG$ are known to be present in 
the diffuse interstellar media (ISM) of spiral galaxies \citep{Beck16}.
Such fields can be loosely divided into large-scale (also known as regular) fields,
which are coherent on scales larger than those of turbulence,
and small-scale (also known as fluctuating) fields, 
which have coherence lengths of at most the outer scale of turbulence, $l$.
A typical estimate for $l$ within the disc of a galaxy is $100\pc$ 
\defcitealias{Ruzmaikin+88}{RSS}\citep[][hereafter \citetalias{Ruzmaikin+88}]{Ruzmaikin+88}.

In the presence of cosmic ray electrons, 
the large-scale magnetic fields of galaxies are sources of polarized synchrotron emission.
The component of the field parallel to the direction of propagation of such emission
causes the plane of polarization to rotate via the Faraday effect.
Likewise, polarized emission from background sources such as active galactic nuclei (AGN) 
will undergo Faraday rotation
as it passes through intervening galaxies on its way to the observer.
Therefore, modelling the large-scale magnetic fields of galaxies may be useful for
predicting and interpreting cosmological radio polarization and RM observations \citep[e.g.][]{Farnes+14}.
Realistic galactic magnetic field models could also be useful 
for helping to constrain the environments of fast radio bursts (FRBs),
using their observed RMs \citep[e.g.][]{Masui+15},
or modelling the transport of cosmic rays of extra-Galactic origin 
from within their source galaxies.

At present there is no analytical solution for galactic large-scale magnetic fields
that is sufficiently generic, realistic, and versatile to be suitable 
for the kinds of applications outlined above.
Phenomenological models \citep{Ferriere+Terral14} are undoubtedly useful for some problems,
but models that are more physically motivated, yet still consistent with observations, are needed.
One approach to painting magnetic fields onto galaxies
is to evolve each galactic field separately using a mean-field dynamo simulation.
Such simulations would use as input evolving parameters from, e.g., 
a semi-analytical galaxy formation model, \citep{Rodrigues+16}.
However, this approach is computationally rather demanding.

An alternative, albeit less rigorous approach, 
is to assume that such dynamo processes lead to saturation 
on timescales small compared with galaxy evolution timescales,
allowing one to adopt a steady-state solution for the magnetic field 
using as input coeval galactic parameter values.
An analytical solution \defcitealias{Chamandy+14b}{CSSS}\citep[][hereafter \citetalias{Chamandy+14b}]{Chamandy+14b} 
has indeed been used for just this purpose \citep{Rodrigues+15}, 
but this solution contains no information about the vertical distribution of magnetic field within the galaxy.
On the other hand, the 3-dimensional, or at least `2.5-dimensional' 
(2.5D, cylindrical symmetry),
spatial structure of the field is required for some of the applications mentioned above.
This motivates the main purpose of the present work: to show how such a 2.5D solution 
of the suitably approximated dynamo equations can be constructed, 
and to demonstrate the efficacy of this solution by comparing it with numerical solutions
of a less approximate set of dynamo equations.

Our model leads to a steady quadrupole-like configuration 
for the regular magnetic field in the saturated state, 
which is consistent with results from standard galactic dynamo theory.
Dynamos in thin accretion discs are probably more complicated, 
with, e.g., oscillatory solutions often obtained 
\citep{Brandenburg+Subramanian05a,Gressel+Pessah15,Moss+16},
and our model is not meant to be applied in such a context.
On the other hand it is not meant to describe any galaxy in particular.
We realize, however, that modelling the field of our own Galaxy,
in particular
\citep[e.g.][]{Vaneck+11,Pshirkov+11,Jansson+Farrar12a,Jansson+Farrar12b},
is important for many investigations, 
such as modelling cosmic ray propagation
or subtracting foreground emission in cosmological studies.
In this context, our model can perhaps serve as a step
toward more physically-motivated Galaxy models.

The paper is organized as follows.
In Section~\ref{sec:model}, we motivate the full set of dynamo equations,
discuss the numerical method used to solve them, 
and summarize the model for the underlying galaxy.
This is followed by an explanation of the analytical method in Section~\ref{sec:analytic}.
Our main results are presented in Section~\ref{sec:results}.
Here we compare analytical and numerical solutions for four different parameter regimes.
Further, for each parameter regime, we compare a suite of numerical solutions,
each obtained from a successively more approximate set of equations.
Thus, a secondary aim is to shed light on the applicability of various approximations used in the literature.
The implications of our results are discussed in Section~\ref{sec:discussion},
and we offer some conclusions in Section~\ref{sec:conclusions}.

\section{Model}
\label{sec:model}
\subsection{Mean-field dynamo theory}
Following the standard prescription, 
the magnetic field $\meanv{B}+\bfb$ and velocity field $\meanv{U}+\bfu$ 
are each written as the sum of a mean (denoted by uppercase) and a fluctuating (denoted by lowercase) component.
Averaging the induction equation we then obtain the standard result
\begin{equation}
  \label{dBdt}
  \frac{\del\meanv{B}}{\del t}=\bfDel\cro\left(\meanv{U}\cro\meanv{B}+\Emf\right),
\end{equation}
where we have neglected terms involving the microscopic (Ohmic) diffusivity since the magnetic Reynolds number $R\magn\gg1$ in galaxies.
Here $\Emf\equiv\overline{\bfu\cro\bfb}$ is the mean electromotive force,
where `bar' denotes mean.
For isotropic turbulence, $\Emf$ can be approximated by the expression \citep{Moffatt78,Krause+Radler80}
\begin{equation}
  \label{Emf}
  \Emf=\alpha\meanv{B} -\eta\bfDel\cro\meanv{B}
\end{equation}
where $\alpha$ can be written as the sum of kinetic and magnetic contributions,
\begin{equation}
  \label{alpha}
  \alpha=\alpha\kin+\alpha\magn,
\end{equation}
with $\alpha\kin=-\tfrac{1}{3}\tau\overline{\bfu\cdot\bfDel\cro\bfu}$ and $\alpha\magn=\tfrac{1}{3}\tau\overline{\bfu\A\cdot\bfDel\cro\bfu\A}$.
Here $\tau$ is the correlation time of the fluctuating flow and $\bfu\A\equiv\bfb/\sqrt{4\pi\rho}$, with $\rho$ the density.
Other terms in equation~\eqref{Emf} may not always be negligible \citep[e.g.][]{Brandenburg+Subramanian05a},
but as our main aim is to present a basic analytical solution,
we leave such complications for future work.
The turbulent magnetic diffusivity $\eta$ is estimated as
\begin{equation}
  \label{beta}
  \eta=\frac{1}{3}\tau u^2.
\end{equation}
The kinetic term $\alpha\kin$, meanwhile, is estimated as \citepalias{Ruzmaikin+88}
\begin{equation}
  \label{alpha_0}
  \alpha\kin= \alpha\f\sin\left(\frac{\pi z}{h}\right)
\end{equation}
with
\begin{equation}
  \label{alpha_0}
  \alpha\f=
  \begin{dcases}
    \frac{\tau^2u^2\Omega}{h},
      & \mbox{if } \Omega\tau\le1;\\
    \frac{\tau u^2       }{h},
      & \mbox{if } \Omega\tau>1,
  \end{dcases}
\end{equation}
where $\Omega$ is the angular velocity.
In our models $\tau u<h$, so $\alpha\kin<u$ \citepalias{Ruzmaikin+88}.
The expression for $\Omega\tau\le1$ is the standard formula of Krause \citep{Krause+Radler80},
while the expression for $\Omega\tau>1$ includes the effects of `rotational saturation' \citepalias[][p.~163]{Ruzmaikin+88}.
For $\Omega\tau\gg1$, $\alpha$ and $\eta$ would be rotationally quenched \citep{Brandenburg+Subramanian05a},
but that case does not arise in our models.

The evolution of $\alpha\magn$ is governed by the dynamical quenching equation \citep{Shukurov+06},
\begin{equation}
  \label{dalpha_mdt}
  \frac{\del\alpha\magn}{\del t}=
   -\frac{2\eta\Emf\cdot\meanv{B}}{l^2B\eq^2}
   -\bfDel\cdot\Flux,
\end{equation}
where 
\begin{equation}
  \label{Beq}
  B\eq= u\sqrt{4\pi\rho}
\end{equation}
is the equipartition field strength.
We have neglected an Ohmic term in equation~\eqref{dalpha_mdt}, 
which is anyway negligible for realistic values of the $\alpha\magn$-flux density $\Flux$.
We assume for simplicity that $l=\tau u$; that is, we assume a Strouhal number $\Strouhal\equiv l/(\tau u)=1$.
The strength of the mean magnetic field in the saturated state is approximately proportional to $l$ 
and thus to $\Strouhal$ \citepalias{Chamandy+14b}.

In general a flux density of the form
\begin{equation}
  \label{flux}
  \Flux=\Flux^\Adv+\Flux^\Diff
\end{equation}
is considered,
where the advective flux density is given by \citep{Subramanian+Brandenburg06},
\begin{equation*}
  \Flux^\Adv=\meanv{U}\alpha\magn
\end{equation*}
and the diffusive flux density by \citep{Brandenburg+09},
\begin{equation*}
  \Flux^\Diff=-\kappa\bmDel\alpha\magn, 
\end{equation*}
with $\kappa$ the turbulent diffusivity of $\alpha\magn$.
Exploring the influence of other potentially important contributions to the helicity flux 
\citep{Vishniac+Cho01,Subramanian+Brandenburg06,Sur+07,Vishniac12b,Vishniac+Shapovalov14,Ebrahimi+Bhattacharjee14} 
is left for future work.
We do, however, compare our results with results using the simple algebraic $\alpha$-quenching formalism,
\begin{equation}
  \label{algebraic}
  \alpha=\frac{\alpha\kin}{1+a(\mean{B}/B\eq)^2},
\end{equation}
with $a$ a parameter which in the literature has typically been set to unity.
We make use of a generalized algebraic quenching formalism
that allows $a$ to be estimated analytically from the dynamical quenching equation~\eqref{dalpha_mdt}.
It is first convenient to define the following dimensionless parameters:
\begin{equation}
  \label{dimensionless}
  \begin{split}
    q\equiv-\frac{\del\ln\Omega}{\del\ln r}, \quad
    H\equiv \frac{h}{\tau u}, \quad
    \Coriolis\equiv\Omega\tau, \quad
    V\equiv \frac{U\f}{u}, \quad
    R_\kappa\equiv\frac{\kappa}{\eta},
  \end{split}
\end{equation}
where $h$ is the density scale height (disc semi-thickness) and $U\f$ is the vertical mean velocity at $z=h$.
In words, $q$ is the radial shear parameter, $H$ is the dimensionless scale-height, $V$ is the dimensionless vertical mean velocity,
and $R_\kappa$ is the ratio of turbulent diffusivities of $\alpha\magn$ and $\meanv{B}$.
The parameter $a$ in the generalized algebraic quenching equation~\eqref{algebraic} is then estimated as \citepalias{Chamandy+14b}
\begin{equation}
  \label{a_quench}
  a=\frac{H^2}{\pi^2R_\kappa +3HV}.
\end{equation}

\subsection{Formalism and equations solved}
As we are dealing with axially symmetric magnetic fields, 
it is convenient to express the magnetic field in terms of scalar potentials $\psi$ and $T$.
Here, the flux function $\psi$ enters through the poloidal field, 
$\meanv{B}\pol\equiv(1/r)\bmDel\psi\cro\hat{\phi}$, and $T\equiv r\mbp$ for the toroidal potential.
Then 
\begin{equation*}
  \mbr=-\frac{1}{r}\frac{\del\psi}{\del z}, \quad \mbp=\frac{T}{r}, \quad \mbz=\frac{1}{r}\frac{\del\psi}{\del r}.
\end{equation*}
(Alternatively, we could have used the variables $\mbp$ and $\mean{A}_\phi=\psi/r$,
where $\meanv{A}$ is the mean vector potential \citep[e.g.][]{Brandenburg+92,Moss+Shukurov01}.)
Here and below, cylindrical polar coordinates ($r$, $\phi$, $z$) are used,
with the galactic angular velocity $\bfOmega$ along the $\bfz$-direction,
and $z=0$ at the galactic midplane.

For simplicity, we assume that mean radial velocities vanish,
that $\Omega$ is independent of $z$,
and that turbulent diffusivities are constant,
but the general equations are provided in Appendix~\ref{sec:full_equations}.
The toroidal and poloidal parts of equation~\eqref{dBdt} and equation~\eqref{dalpha_mdt} can then respectively be written as
\begin{equation}
  \begin{split}
    \label{dTdt}
    \frac{\del T}{\del t}=&
      -\frac{\del}{\del z}(\muz T)
      +q\Omega\frac{\del\psi}{\del z}-\alpha\Lambda^-\psi  \\& 
      -\frac{\del\alpha}{\del r}\frac{\del\psi}{\del r} -\frac{\del\alpha}{\del z}\frac{\del\psi}{\del z} +\eta\Lambda^-T,
  \end{split}
\end{equation}
\begin{equation}
  \label{dpsidt}
  \frac{\del\psi}{\del t}= -\muz\frac{\del\psi}{\del z} +\alpha T +\eta\Lambda^-\psi,
\end{equation}
\begin{equation}
  \begin{split}
    \label{dalpha_mdt_formalism}
    \frac{\del\alpha\magn}{\del t}=&
      -\frac{2\eta}{l^2r^2B\eq^2}\Biggl\{\alpha\left[\left(\frac{\del\psi}{\del r}\right)^2 
                                               +T^2 +\left(\frac{\del\psi}{\del z}\right)^2\right]\Biggr.\\&
      \Biggl.-\eta\left[\frac{\del\psi}{\del r}\frac{\del T}{\del r} 
                 -T\Lambda^-\psi +\frac{\del\psi}{\del z}\frac{\del T}{\del z}\right]\Biggr\}
      -\frac{\del}{\del z}(\muz\alpha\magn)\\&
      +\kappa\Lambda^+\alpha\magn,
  \end{split}
\end{equation}
where $\Lambda^\pm\equiv \del^2/\del r^2 \pm(1/r)\del/\del r +\del^2/\del z^2$,
and $\alpha$ is generally given by equation~\eqref{alpha}, 
but sometimes by equation~\eqref{algebraic}.

\subsection{Boundary conditions and numerical setup}
\label{sec:boundary}
Horizontal and vertical boundary conditions must be chosen for the variables $\psi$, $T$ and $\alpha\magn$.
Evidently, $T=r\mbp=0$ at $r=0$.
We further demand that, by symmetry, $\mbp(t,0,z)\rightarrow0$ as $r\rightarrow0$; 
this then implies $\del T/\del r=0$ at $r=0$.
Finiteness of $\mbr(t,0,z)$ implies $\del\psi/\del z|_{r=0}=0$; 
we choose $\psi(t,0,z)=0$ without loss of generality.
This is the natural choice since $\psi(t,r,z)$ is then proportional 
to the magnetic flux through a horizontal disc of radius $r$ centred at the position $(0,z)$,
\begin{equation}
  \psi(t,r,z)\propto \displaystyle\int_0^r\mbz(t,r',z) r' dr'.
\end{equation}
Finiteness of $\mbz(t,0,z)$ implies $\del\psi/\del r|_{r=0}=0$.
Note that $\mbr\rightarrow0$, while $\mbz\rightarrow\del^2\psi/\del r^2|_{r=0}$, as $r\rightarrow0$,
and these values of $\mbr|_{r=0}$ and $\mbz|_{r=0}$ are set explicitly.
We also set $\del\alp\magn/\del r|_{r=0}=0$ to avoid a singularity in the diffusive flux term at the origin.
Further, we adopt the same boundary conditions $\psi=T=\del\psi/\del r=\del T/\del r=\del\alpha\magn/\del r=0$ at $r=R$, 
corresponding to the outermost radius of the simulation domain;
solutions are insensitive to the choice of boundary conditions at $r=R$ 
since $R$ is chosen to be well outside the region of dynamo action.
We set $R=15\kpc$.

We impose vacuum boundary conditions $\mbr=\mbp=0$ at $z=\pm h$ \citepalias{Ruzmaikin+88}, 
which implies $\del\psi/\del z=T=0$ at $z=\pm h$.
Thus, the mean magnetic field is forced to be vertical outside the disc.
In imposing these boundary conditions we ignore the gaseous halo, 
leaving its inclusion for future work.
To see how these boundary conditions arise,
begin with the requirement that $\bfDel\cro\meanv{B}=0$ outside the disc. 
This implies $\del T/\del z=0$. 
Obviously, $T\rightarrow0$ as $|z|\rightarrow\infty$,
which then implies $T=0$ at $z=\pm h$.
Setting $\bfDel\cro\meanv{B}=0$ also leads to the condition
$\del^2\psi/\del z^2=-\del^2\psi/\del r^2+(1/r)\del\psi/\del r$. 
The right-hand side is usually small compared with the left hand side, 
so this condition can be approximated by $\del^2\psi/\del z^2=0$ outside the disc.
This approximation becomes invalid for $r\lesssim h$, so our model is less reliable at the very centre of the disc.
Since $\mbr\rightarrow0$ as $|z|\rightarrow\infty$ we must have $\del\psi/\del z=0$ at $z=\pm h$.
Further, we set $\del^2\alp\magn/\del z^2=0$ at $z=\pm h$, which allows $\alpha\magn$ to flow across the disc boundary.

We have also experimented with other choices of boundary conditions on $z=\pm h$.
We tried various combinations of $\del\alpha\magn/\del z=0$ (instead of $\del^2\alpha\magn/\del z^2=0$),
$\del^2\psi/\del z^2=0$ so that $\del\mbr/\del z=0$ (instead of $\del\psi/\del z=0$) 
and $\del T/\del z=0$ so that $\del\mbp/\del z=0$ (instead of $T=0$) at $z=\pm h$.
For example, choosing $\del^2\psi/\del z^2=\del T/\del z=0$ implies $\del\mbr/\del z=\del\mbp/\del z=0$.
Interestingly, for models with a strong outflow,
changing the boundary conditions in any of these ways has almost no effect on the solution.
However, for models without an outflow (or with a weak outflow),
solutions are rather sensitive to the vertical boundary conditions.
The explanation for this is that if the outflow removes the field and $\alpha\magn$ at the boundaries rapidly enough,
boundary effects cannot propagate inward.
Changing the boundary condition on $\alpha\magn$ to $\del\alpha\magn/\del z=0$ at $z=\pm h$ 
has little effect on the solutions for $\meanv{B}$, even if no outflow is present.
However, if $\del^2\psi/\del z^2=T=0$ is adopted at the vertical boundaries,
then $\alpha\magn$ undergoes a sign change in boundary layers near the disc surface,
leading to magnetic field configurations that are rather different from the standard solutions
(for models without a strong outflow).
On the other hand, if we choose $\del T/\del z=0$ instead of $T=0$ at the boundaries,
we find that either the code does not converge or, 
if $\del^2\psi/\del z^2=0$ at $z=\pm0$ is also chosen,
the field decays.

We use an $r$--$z$ grid that is linear in both coordinates,
and solve the equations using the same finite differencing (6th order)
and Runge-Kutta time-stepping (3rd order) routines 
employed in the Pencil Code \citep{Brandenburg03}.
For thin disc solutions, we use a resolution $N_r\times N_z=801\times101$, 
excluding ghost zones, for the runs presented,
while for the model with a thicker disc, we use $N_r\times N_z=801\times201$.
A much smaller resolution, e.g. $41\times41$, is usually sufficient for testing purposes.
Runs with even higher resolution were performed as a check for convergence, when necessary.

Initial conditions at time $t=0$ are $\psi=T=10^{-4}(r/R)^2(1-r/R)\Exp{-r/R}\cos^2[\pi z/(2h)]\Exp{-z^2/h^2}$ 
and $\alpha\magn=0$.
The form of the seed field is chosen so as to be relatively simple and to satisfy the boundary conditions,
though solutions do not depend on initial conditions as long as the seed field is sufficiently small.

Although our dynamo model includes the Lorentz force via the dynamical quenching non-linearity,
we assume that the dynamo parameters (with the exception of $\alpha\magn$) 
do not evolve with time, e.g. in response to the evolution of the mean magnetic field.
As we are mainly concerned with steady-state solutions in the saturated regime, 
this is not an important limitation.
Below we describe the model of the underlying galaxy within which the dynamo operates.

\subsection{Galaxy model}
\label{sec:galaxy_model} 
The mean velocity field takes the form ($0$, $\mup$ ,$\muz$) in cylindrical coordinates,
with $\mup=r\Omega$.
We employ a Brandt rotation curve of the form
\begin{equation}
  \Omega(r)= \frac{\mup(8\kpc)}{8\kpc}\sqrt{\frac{1+(8\kpc/r_\Omega)^2}{1+(r/r_\Omega)^2}},
\end{equation}
where $\mup(8\kpc)=220\kms$ and $r_\Omega=2\kpc$ are parameters.
Note that $\Omega$ is a maximum at $r=0$, 
and that $\mup$ approaches a constant value as $r\rightarrow\infty$.
The vertical velocity is given by
\begin{equation}
  \muz(z)=U\f\frac{z}{h},
\end{equation}
with the amplitude $U\f$ a parameter which is equal to the magnitude of the vertical velocity at the disc boundary.

We adopt an exponential profile for the equipartition energy,
\begin{equation}
  B\eq^2(r,z)=B\f^2\Exp{(-r/r\f-|z|/h)},
\end{equation}
where $B\f$ is the equipartition field strength at $r=z=0$,
and we set the radial scale length $r\f=7.5\kpc$ \citep[e.g.][]{Beck07}.
Below, magnetic fields are quoted in the arbitrary unit $B\f$.

Table~\ref{tab:models} lists the various parameter sets (`models') explored in this work.
For simplicity, we adopt constant values of $h$, $l$, $u$, and $U\f$.
All models employ a disc with constant turbulent velocity $u=12\kms$.
Model~A, with $h=0.5\kpc$, $l=0.1\kpc$ and $U\f=0$, is the fiducial model.
Model~B differs from Model~A in having a nonzero outflow velocity $U\f=3\kms$,
Model~C differs from Model~A in having a twice larger value of the turbulent scale $l=0.2\kpc$,
and thus also of $\tau=l/u$,
while Model~D differs from Model~A 
in having a twice larger value of the disc half-thickness $h=1\kpc$.

\begin{table}
  \begin{center}
    \caption{Parameter values for our four models.
             All models employ a disc of constant turbulent speed $u=12\kms$.
             The fiducial model, Model~A, lacks an outflow and assumes a turbulent outer scale $l=\tau u=0.1\kpc$
             and disc semi-thickness $h=0.5\kpc$.
             Model~B differs from the fiducial model by having a nonzero vertical outflow speed $U\f=3\kms$ 
             at the disc surface.
             Model~C has no outflow, but assumes $l=\tau u=0.2\kpc$,
             while Model~D is like Model~A but uses $h=1\kpc$.
             Dimensionless parameters $H$, $V$ and $\Coriolis|_{r=0}$ defined in equations~\eqref{dimensionless}
             are also provided.
    }
    \label{tab:models}
    \begin{tabular}{@{}lllllll@{}}
      \hline
      Model     &$h$        &$l$          &$U\f$          &$H$    &$V$    &$\Coriolis(0,z)$  \\
                &$[\!\kpc]$ &$[\!\kpc]$   &$[\!\kms]$     &       &       &             \\
      \hline                            
      A         &$0.5$      &$0.1$        &$0$            &$5$    &$0$    &$0.94$       \\
      B         &$0.5$      &$0.1$        &$3$            &$5$    &$1/4$  &$0.94$       \\
      C         &$0.5$      &$0.2$        &$0$            &$5/2$  &$0$    &$1.89$       \\
      D         &$1$        &$0.1$        &$0$            &$10$   &$0$    &$0.94$       \\
      \hline
    \end{tabular}
  \end{center}
\end{table}

\section{Analytical method}
\label{sec:analytic}
The main purpose of the present work is to present a 2.5D (axisymmetric)
analytical model that can be used for predicting observations,
and to show that it compares favourably with full numerical solutions.
To build such a model we combine two types of analytical solution useful for mean-field galactic disc dynamos.
We refer to the first type as the `local no-$z$ solution', 
or, in the final steady-state after non-linear saturation takes place, the `local critical no-$z$ solution.'
Here, we neglect radial derivatives (the local or slab approximation, \citetalias{Ruzmaikin+88}), 
replace $z$-derivatives by divisions by the scale-height $h$, with suitable numerical coefficients 
(the no-$z$ approximation, \defcitealias{Chamandy+Taylor15}{CT} 
\citealt{Subramanian+Mestel93,Moss95,Phillips01}; \citetalias{Chamandy+14b};
\citealt[][hereafter \citetalias{Chamandy+Taylor15}]{Chamandy+Taylor15}),
and finally, assume that $\alpha$ takes on its critical value $\alpha\crit$ in the saturated state.
The latter assumption is quite natural since the critical value 
is defined to be that which gives vanishing growth rate of the mean magnetic field.

The second type of useful analytical solution is the `local perturbation solution' 
(\citetalias{Ruzmaikin+88}; \citealt{Ji+14}; \citetalias{Chamandy+14b}), 
or, for the saturated regime, the `local critical perturbation solution.'
Here the terms in the local (slab) mean-field induction equation that involve the mean velocity field and dynamo source terms
are treated as perturbations to the mean-field diffusion equation.
Solutions depending on $z$, and depending parametrically on $r$, 
are obtained to a specified order in the dimensionless quantities $\sqrt{-D\crit}$ and $R_U=hU\f/\eta$,
which are measures of the $\alpha\Omega$ dynamo action in the saturated state 
and strength of the vertical outflow from the disc, respectively.
More generally, the dynamo number $D=-\alpha q\Omega h^3/\eta^2$,
so that its critical value $D\crit=-\alpha\crit q\Omega h^3/\eta^2$,
while we denote its value in the kinematic regime by  $D\f=-\alpha\f q\Omega h^3/\eta^2$.
In what follows it is convenient to parameterize equations by $\Dtilde\equiv D\f/D\crit=\alpha\f/\alpha\crit$.

Below we summarize the key results of each type of solution and explain how they can be combined.
The reader is referred to \citetalias{Chamandy+14b} and \citetalias{Chamandy+Taylor15} for further details.

\begin{table*}
    \caption{Summary of the various levels of approximation presented.
             The bottom-most three approximations are combined 
             to construct the 2.5D analytical solution.
    }
    \label{tab:approximations}
     \begin{minipage}{1\textwidth}
     \begin{center}
     \begin{tabular}{@{}lp{10cm}l@{}}
      \hline
      Approximation            &Description                                                    
     &Relevant figures   \\
      \hline              
      \hline
      $\alpha$-$\Omega$        &Terms involving $\alpha$ neglected in $\phi$-component of equation~\eqref{dBdt}.
     &Panels (b)--(f) of Figures~\ref{fig:Bicontour_A}--\ref{fig:Bicontour_D}.   \\
      \hline
      Thin disc                &Terms involving $\mbz$ neglected;\footnote{Except for terms involving $\del\mbz/\del z$; 
                                                                           see Section~\ref{sec:approx}.} 
                                $\mbz$ computed using $\bfDel\cdot\meanv{B}=0$.
     &Panels (c)--(f) of Figures~\ref{fig:Bicontour_A}--\ref{fig:Bicontour_D}.                      \\
      \hline
      Slab                     &Terms involving radial derivatives neglected (encompasses thin disc approximation).
     &Panels (d)--(f) of Figures~\ref{fig:Bicontour_A}--\ref{fig:Bicontour_D}.                      \\
      \hline
      Algebraic quenching      &Approximates the non-linearity of equation~\eqref{dalpha_mdt} 
                                by replacing it with expression~\eqref{algebraic} with \eqref{a_quench}.
     &Panels (e) of Figures~\ref{fig:Bicontour_A}--\ref{fig:Bicontour_D}.                      \\
      \hline
      Perturbation             &Terms of equation~\eqref{dBdt} involving $\alpha$ and $\meanv{U}$ 
                                treated as perturbations to solutions of the diffusion equation \citepalias[see][]{Chamandy+14b}.
     &                      \\
      No-$z$                   &Vertical derivatives replaced by divisions by $h$
                                with appropriate numerical coefficients \citepalias[see][]{Chamandy+14b,Chamandy+Taylor15}.  
     &Panels (f) of Figures~\ref{fig:Bicontour_A}--\ref{fig:Bicontour_D}.     \\
      Steady-state (critical)  &Time derivatives set to zero.        
     &                      \\
      \hline
    \end{tabular}
    \end{center}
    \rput[lt](13.4,2.3){$\left.\rule{0cm}{1.0cm}\right\}$}
    \end{minipage}
\end{table*}

\subsection{No-$z$ solution}
\label{sec:noz}
We first summarize the local critical no-$z$ solution.
To begin with, 
the slab mean induction equation reduces to a set of $z$- and $t$-independent algebraic equations,
parameterized by the radius $r$.
These equations can be solved to yield vertical averages for $\mbr$ and $\mbp$,
while $|\mbz|$ can be estimated from the condition $\bfDel\cdot\meanv{B}=0$.
For the kinematic regime at least, the local solutions can be extended to the global domain using WKBJ theory;
we use this method below to estimate the global growth rate of the mean magnetic field.

In terms of the parameters \eqref{dimensionless}, the \textit{local} exponential growth rate in the kinematic regime is given by
\begin{equation}
  \label{gamma}
  \gamma= 
  \begin{dcases}
    \sqrt{\frac{2q}{\pi}}\frac{\Omega}{H}
      \left[1 -\frac{1}{12}\sqrt{\frac{\pi}{2q}}\left(\frac{\pi^2 +6HV}{H\Coriolis}\right)\right],& \mbox{ if } \Coriolis\le1;\\
    \sqrt{\frac{2q}{\pi}}\frac{\Omega}{H}
      \left[\frac{1}{\sqrt{\Coriolis}} 
            -\frac{1}{12}\sqrt{\frac{\pi}{2q}}\left(\frac{\pi^2 +6HV}{H\Coriolis}\right)\right],& \mbox{ if } \Coriolis>1.
  \end{dcases}
\end{equation}
This is the growth rate that obtains, at a given $r$, if radial transport terms are not included.
Another quantity of interest is the magnetic pitch angle $p=\tan^{-1}(\mbr/\mbp)$, with $-\pi/2<p\le\pi/2$.
In the saturated state, $p=p\sat$, with \citepalias{Chamandy+Taylor15}
\begin{equation}
  \label{psat}
  \tan p\sat = -\frac{\pi^2+6H V}{12qH^2\Coriolis},
\end{equation}
and in the kinematic regime $p=p\kin$ with
\begin{equation}
  \label{pkin}
  \tan p\kin= 
  \begin{dcases}
    -\frac{1}{H}\sqrt{\frac{2}{\pi q    }},& \qquad \mbox{ if } \Coriolis\le1;\\
    -\frac{1}{H}\sqrt{\frac{2}{\pi q\Coriolis}},& \qquad \mbox{ if } \Coriolis>1.
  \end{dcases}
\end{equation}

To obtain the magnetic energy density in the saturated state,
we must make use of equation~\eqref{dalpha_mdt} for $\alpha\magn$.
We then obtain \citepalias{Chamandy+14b},
\begin{equation}
  \label{Bsat}
  \mean{B}\sat^2= B\eq^2\frac{(\Dtilde -1)\left(\pi^2 R_\kappa +3HV\right)}{2H^2\xi(p\sat)},
\end{equation}
where 
\begin{equation}
  \xi(p)=1-\frac{3\cos^2p}{4\sqrt{2}},
\end{equation}
and where the normalized dynamo number $\Dtilde=\alpha/\alpha\crit=\tan^2p\kin/\tan^2p\sat$ is given by
\begin{equation}
  \label{Dtilde}
  \Dtilde= 
  \begin{dcases}
    \frac{288q}{\pi    }\left(\frac{H\Coriolis}{\pi^2+6HV}\right)^2,& \qquad \mbox{ if } \Coriolis\le1;\\
    \frac{288q}{\pi\Coriolis}\left(\frac{H\Coriolis}{\pi^2+6HV}\right)^2,& \qquad \mbox{ if } \Coriolis>1.
  \end{dcases}
\end{equation}
Note that $\Dtilde$ must be greater than unity for a supercritical ($\gamma>0$) dynamo.
That is, the dynamo number in the kinematic regime $D\f$
must exceed in magnitude the critical dynamo number, which is given by
\begin{equation}
  \label{Dcnoz}
  D\crit=-\frac{\pi}{32}\left(\pi^2+6HV\right)^2= -\frac{\pi}{2}(3qH^2\Coriolis)^2\tan^2p\sat.
\end{equation}
Note that for $\cos^2p\rightarrow1$, we obtain the convenient result $\xi(p\sat)\simeq1/2$.
We also note in passing that to write down these no-$z$ solutions, 
certain numerical coefficients had to be calibrated to numerical solutions,
but for our purposes these can now be thought of as fixed;
here we use the expressions from \citetalias{Chamandy+Taylor15}.

In the kinematic regime ($\mean{B}^2\ll B\eq^2$) we have $\meanv{B}\propto\exp(\Gamma t)$,
where $\Gamma$ is the growth rate of the fastest growing global eigenmode.
To estimate the global growth rate,
the standard procedure is to separate a suitably simplified version of the mean-field induction equation~\eqref{dBdt}
into two equations: one equation for the local ($z$-dependent) part 
and another for the global ($r$-dependent, or, in general, $r$- and $\phi$-dependent) part \citepalias{Ruzmaikin+88}.
Here we consider the case of a spatially constant turbulent diffusivity $\eta$.
The Schr\"{o}dinger-type global equation that results from this analysis
can then be solved approximately using WKBJ theory \citep[e.g.][]{Chamandy+13b}.
This leads to an eigen condition for the fastest growing mode,
\begin{equation}
  \label{eigen}
  \displaystyle\int^{r_+}_{r_-}[E-W(r)]^{1/2}dr=\frac{\pi}{2},
\end{equation}
where the `potential' is given by
\begin{equation}
  \label{W}
  W(r)=\frac{1}{r^2}-\frac{\gamma}{\eta},
\end{equation}
the `energy' eigenvalue by
\begin{equation}
  \label{E}
  E=-\frac{\Gamma}{\eta},
\end{equation}
and where the integration limits in equation~\eqref{eigen}, $r_-$ and $r_+$, are the zeros of the integrand.
Equation~\eqref{eigen} with equations~\eqref{W} and \eqref{E} is solved by numerical iteration to yield $\Gamma$.
One could also calculate the kinematic eigenfunction under the WKBJ approximation, 
but as we are mainly interested in saturated solutions, we do not attempt this.
Below we summarize the other type of analytical solution needed to construct our hybrid analytical solution:
the local critical perturbation solution.

\subsection{Perturbation solution}
\label{sec:perturbation}
Like the no-$z$ solution, the perturbation solution contains a parameteric dependence on $r$, 
but unlike the no-$z$ solution it retains a dependence on $z$.
However the local critical perturbation solution is not sufficient on its own because it is unnormalized.

In terms of the parameters~\eqref{dimensionless}, solutions for the field components $\mbr$ and $\mbp$ in the saturated state are
\begin{equation}
  \label{Brpert}
  \begin{split}
    \mbr=& K \left(-\frac{D\crit^\mathrm{pert}}{3qH^2\Coriolis}\right)
           \Bigg[
                                                              \cos\left(                \frac{ \pi z}{2h}\right) \\
         & +\frac{3}{4\pi^2}\left(\sqrt{-\pi D\crit^\mathrm{pert}} -\frac{3HV}{2}\right)
                                                              \cos\left(                \frac{3\pi z}{2h}\right) \\
         & +\frac{3HV}{2\pi^2}\displaystyle\sum^\infty_{n=2}\frac{(-1)^n(2n+1)}{n^2(n+1)^2}
                                                              \cos\left(\frac{\left(2n+1\right)\pi z}{2h}\right)
           \Bigg],
  \end{split}                                                 
\end{equation}
and
\begin{equation}
  \label{Bppert}
  \begin{split}
    \mbp=& K \left(-2\sqrt{-\frac{D\crit^\mathrm{pert}}{\pi}}\right)
           \Bigg[
                                                              \cos\left(                \frac{ \pi z}{2h}\right) \\
         & +\frac{3}{4\pi^2}\left(                   -\frac{3HV}{2}\right)
                                                              \cos\left(                \frac{3\pi z}{2h}\right) \\
         & +\frac{3HV}{2\pi^2}\displaystyle\sum^\infty_{n=2}\frac{(-1)^n(2n+1)}{n^2(n+1)^2}
                                                              \cos\left(\frac{\left(2n+1\right)\pi z}{2h}\right)
           \Bigg],
  \end{split}                                                 
\end{equation}
where
\begin{equation}
  \label{Dcpert}
    D\crit^\mathrm{pert}=-\frac{\pi^3}{4}\left\{ \frac{ 1 +6HV/\pi^2 
                                                -( 18H^2V^2/\pi^4)( 1 -\pi^2/6)}{ 1 +9HV/[ 2\pi( \pi +4)]}\right\}^2.
\end{equation}
and $K$ is a parameter which can be estimated using the local critical no-$z$ solution, 
as explained in Section~\ref{sec:combined}.
Note that $D\crit^\mathrm{pert}$ in equation~\eqref{Dcpert} is slightly different from $D\crit$ obtained 
from the no-$z$ approximation in equation~\eqref{Dcnoz}, 
which is why the former has been labelled with a superscript.

\subsection{Combined no-$z$ and perturbation solution}
\label{sec:combined}
What has not been done previously is to combine these two types of analytical solution for $\meanv{B}$ in the saturated state
to obtain a normalized axisymmetric solution that depends on both $r$ (parametrically) and $z$.
To do this, one must use the no-$z$ solution to estimate the parameter $K$,
hence normalizing the perturbation solution.
We simply set
\begin{equation}
  \label{C0}
  K(r)=\pm\frac{\mean{B}\sat(r,0)}{\sqrt{\langle\Btilde^2\rangle}},
\end{equation}
where $\mean{B}\sat$ is obtained from equation~\eqref{Bsat}, and
\begin{equation}
  \langle\Btilde^2\rangle= \frac{1}{2h}\displaystyle\int_{-h}^h(\mbrtilde^2 +\mbptilde^2)dz.
\end{equation}
Here $\mbrtilde$ and $\mbptilde$ are the magnetic field components of expressions~\eqref{Brpert} and \eqref{Bppert} normalized by $K$.
The `$\pm$' in equation~\eqref{C0} comes from the sign invariance of the induction equation,
and the sign must be chosen to match that of the numerical solution, 
which ultimately depends on the arbitrary seed field chosen.
Note also that $\mean{B}\sat$ is proportional to $B\eq$, which varies exponentially with $|z|$ in our model,
so to remove this $z$-dependence, we adopt the midplane value of $\mean{B}\sat$ in equation~\eqref{C0}.
This choice is partly motivated by the good \textit{a posteriori} agreement 
we obtain with numerical solutions (Section~\ref{sec:results}).
Thus, equations \eqref{Brpert} and \eqref{Bppert}, with equations~\eqref{C0}, \eqref{Dcpert}, and \eqref{Bsat} 
comprise the analytical solutions for $\mbr(r,z)$ and $\mbp(r,z)$ in the saturated state.
To estimate $\mbz(r,z)$, we make use of the property $\bfDel\cdot\meanv{B}=0$.
Then, in cylindrical coordinates we have
\begin{equation}
  \mbz(r,z)=\displaystyle-\int_0^z\frac{1}{r}\frac{\del}{\del r}[r\mbr(r,z')]dz',
\end{equation}
where $z'$ is the integration variable, and where we have made use of the fact that for quadrupolar solutions $\mbz(r,0)=0$.

The prescription presented above is rather trivial, as it relies simply on combining, 
in a straightforward way, two known analytical solutions.
Yet it is potentially powerful because it generates a magnetic field solution that is effectively global and, 
though it is 2.5D, can be utilized in a 3D context.
Below we present examples of these analytical solutions and compare them with numerical solutions
under various levels of approximation.
The different types of approximation are summarized in Table~\ref{tab:approximations},
and discussed further below.

\begin{figure*}                     
  \includegraphics[width=58mm,clip=true,trim=  10 10 10 10]{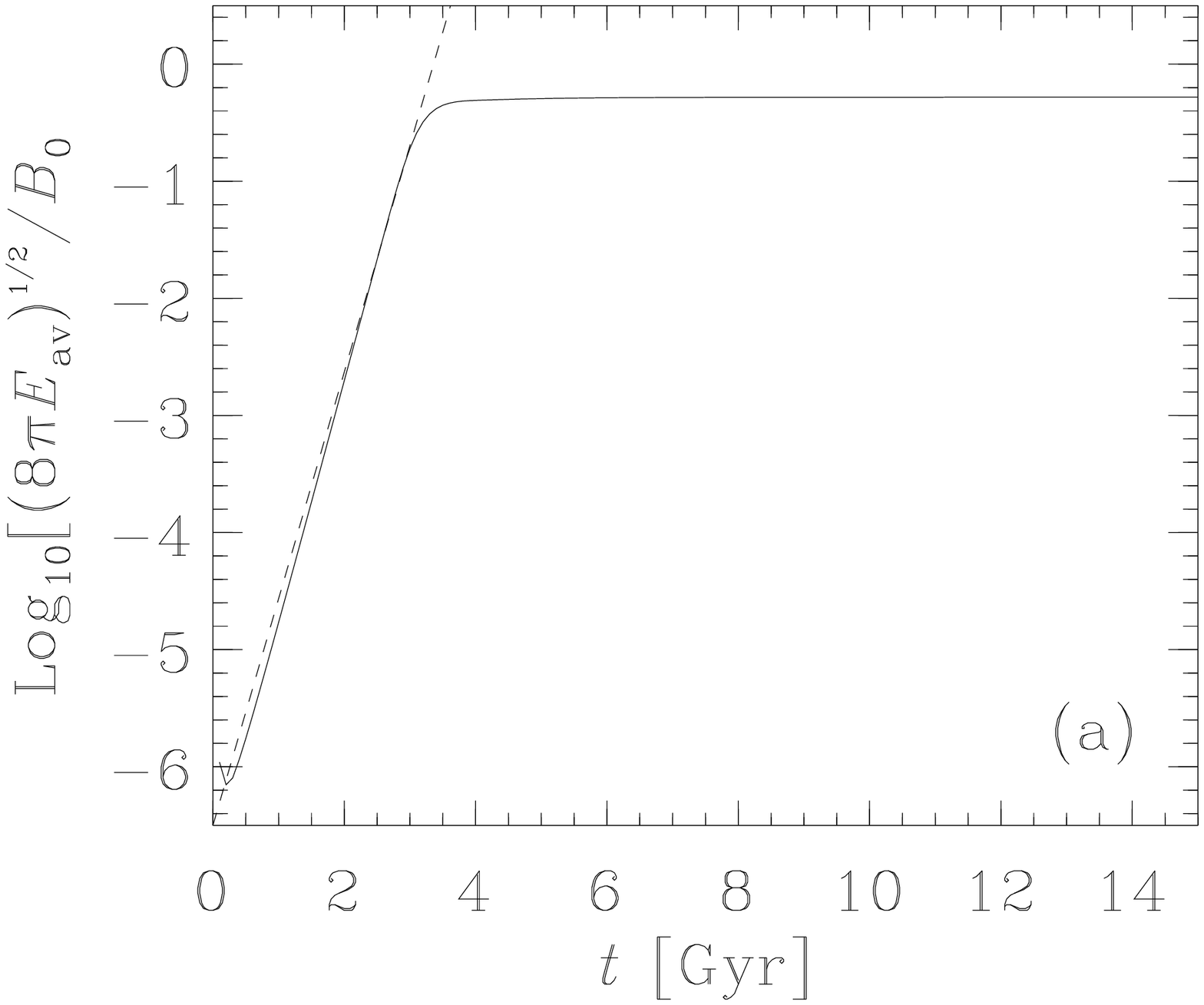}
  \includegraphics[width=58mm,clip=true,trim=  10 10 10 10]{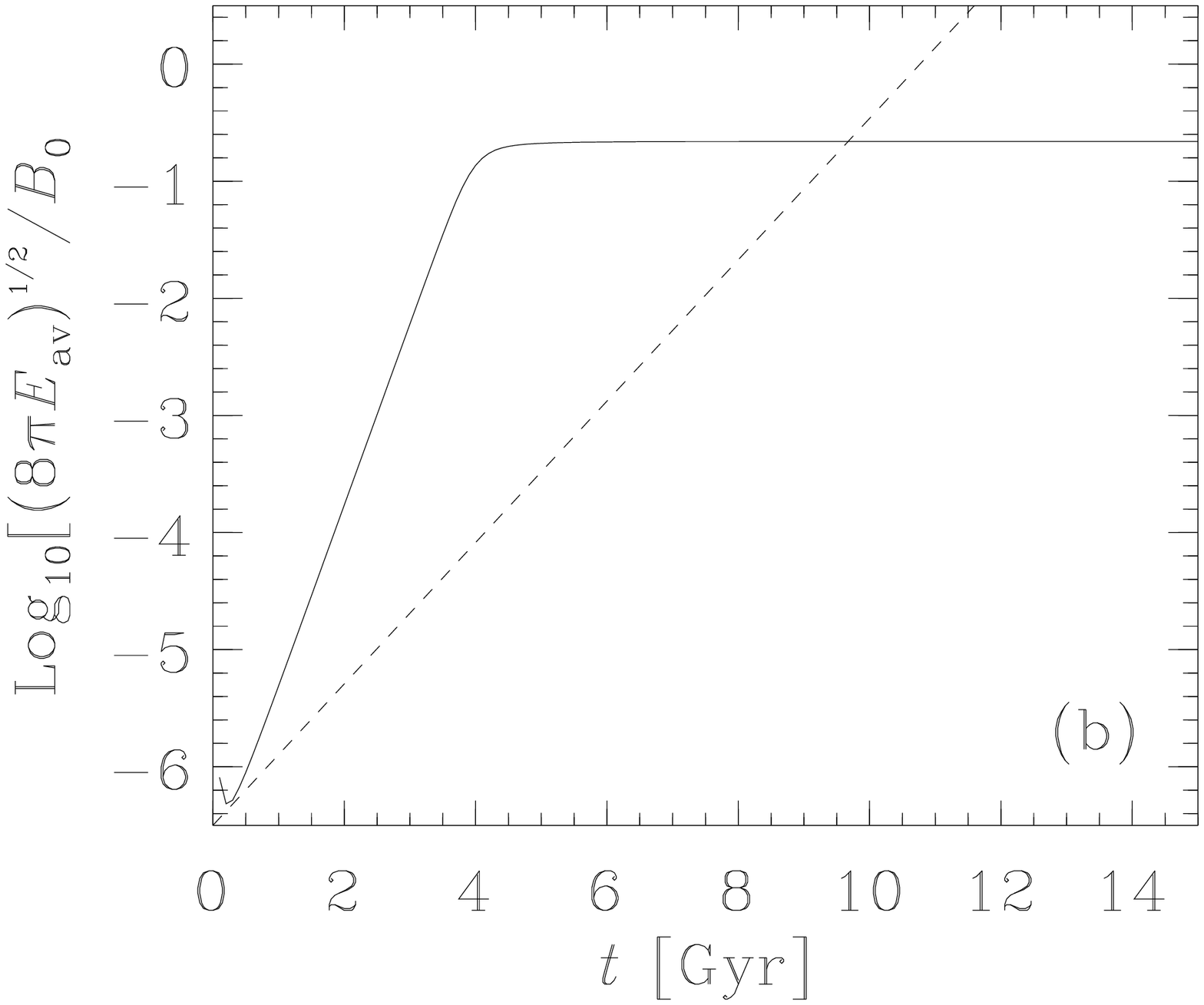}
  \includegraphics[width=58mm,clip=true,trim=  10 10 10 10]{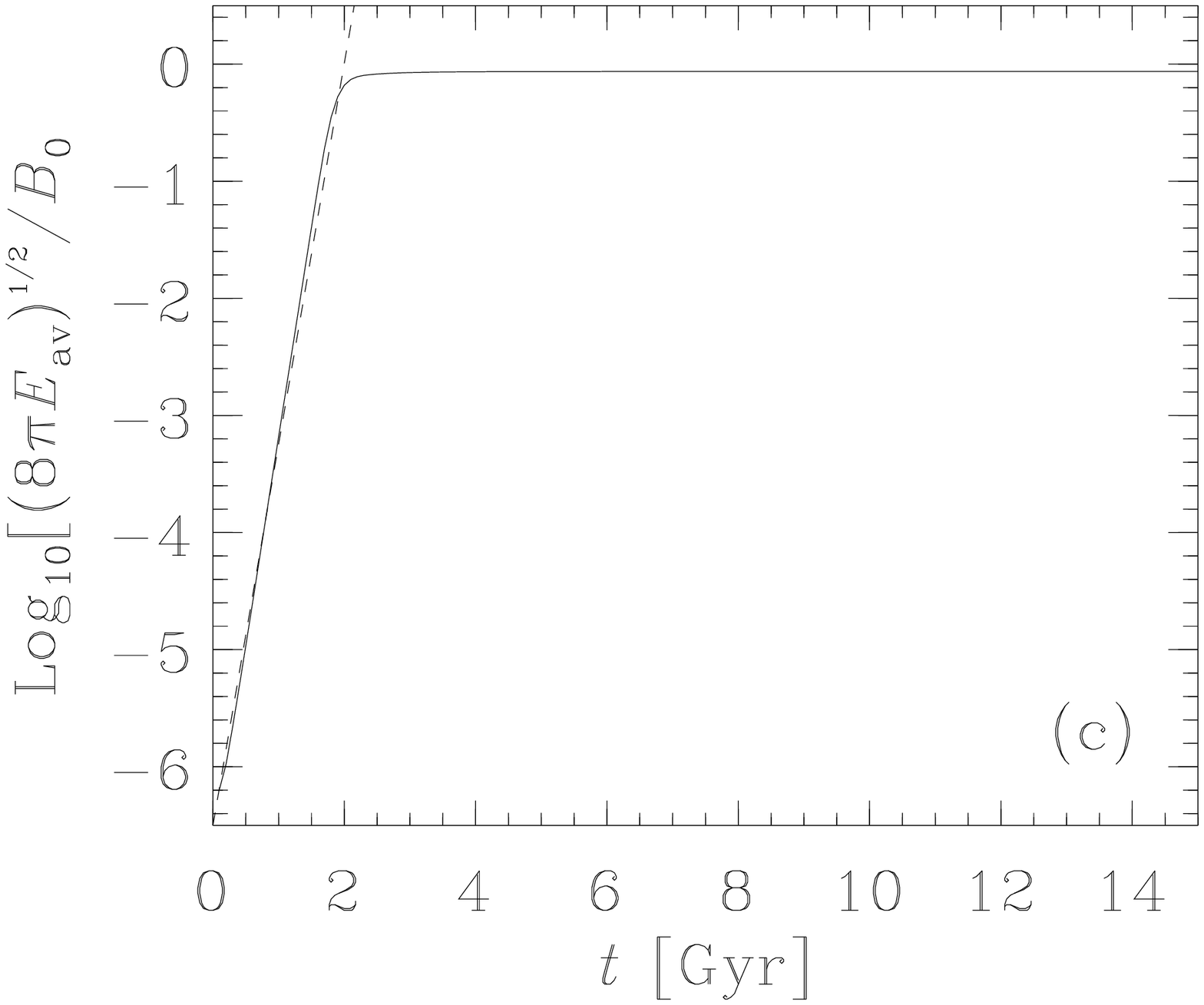}\\
  \includegraphics[width=58mm,clip=true,trim=  10 10 10 10]{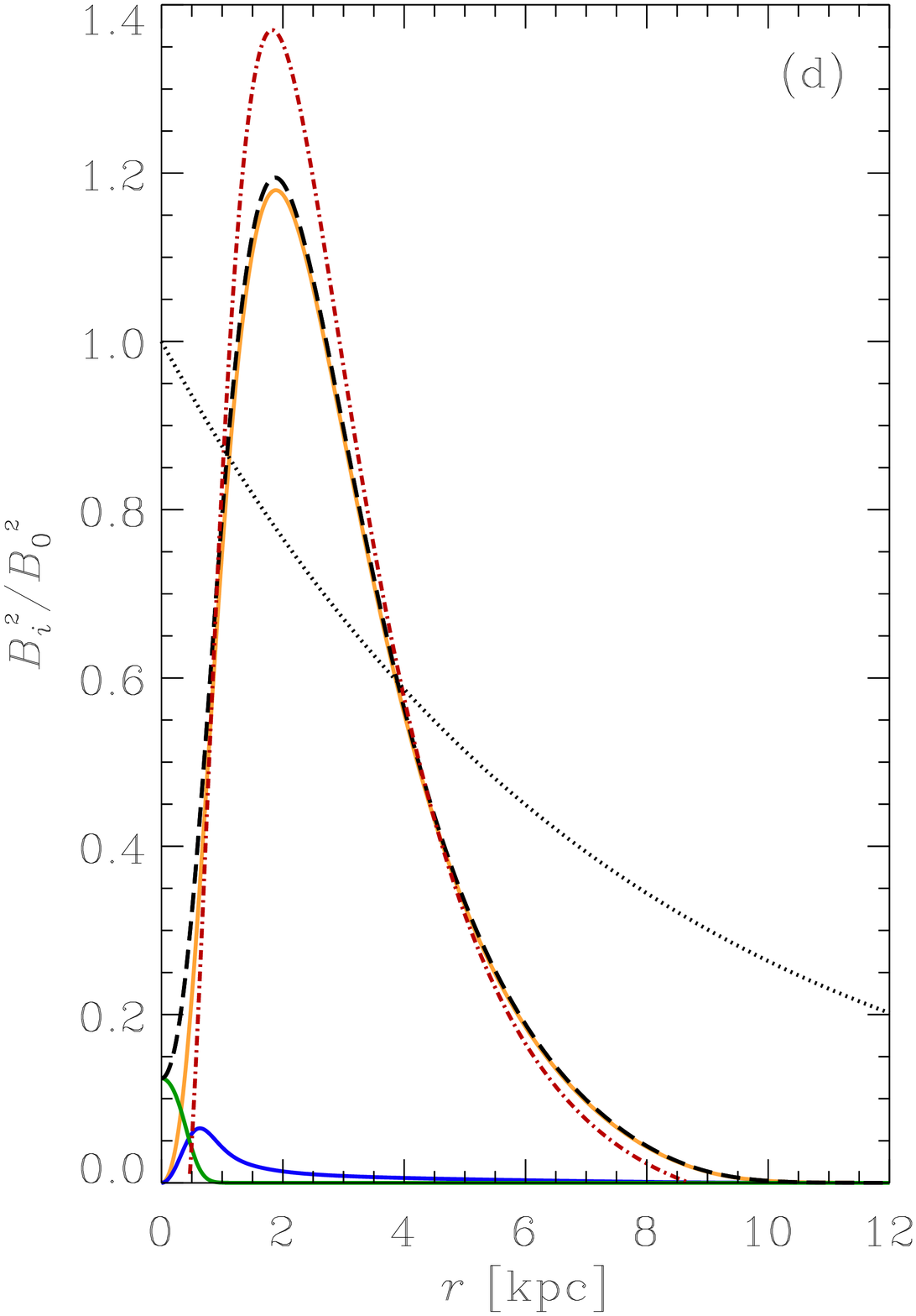}
  \includegraphics[width=58mm,clip=true,trim=  10 10 10 10]{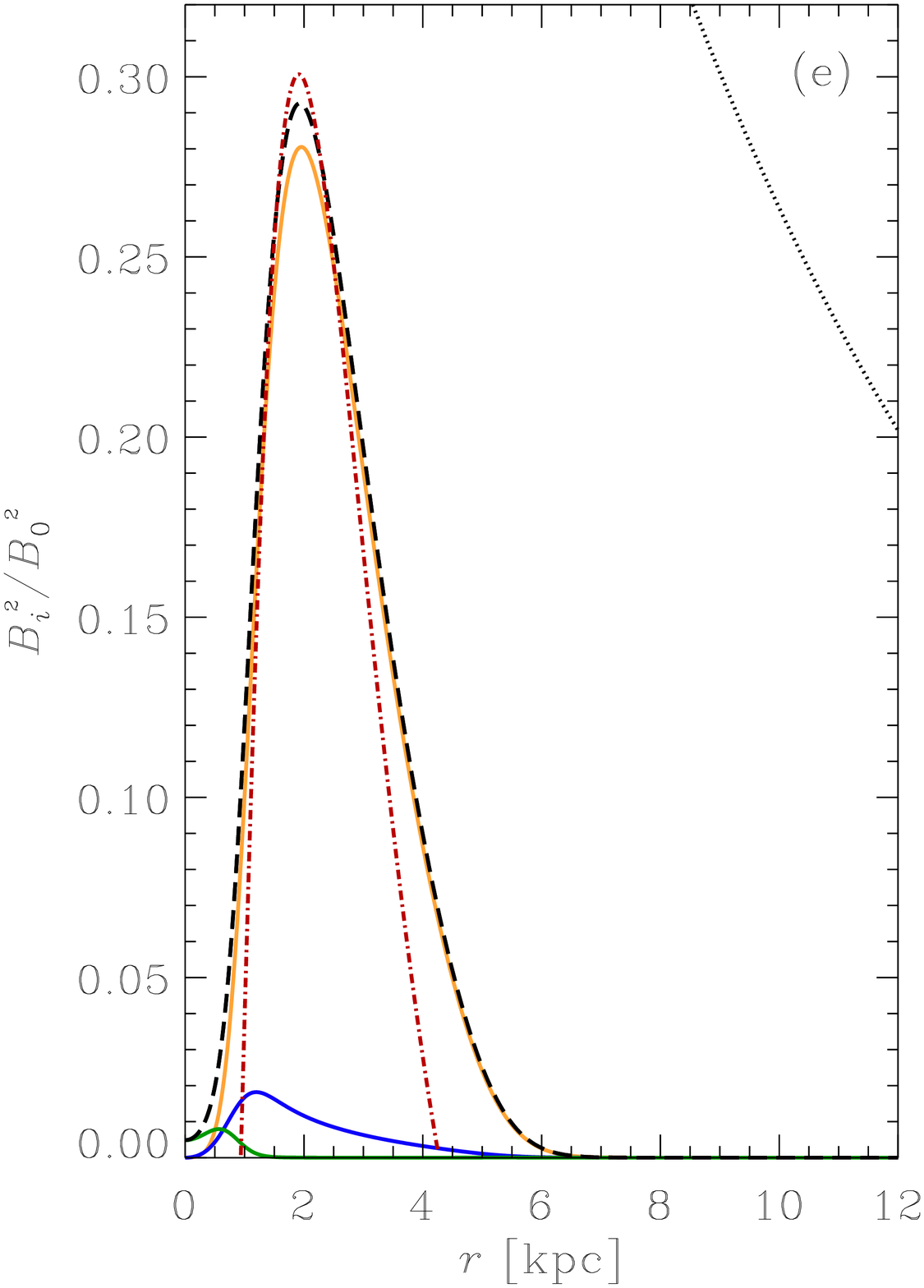}
  \includegraphics[width=58mm,clip=true,trim=  10 10 10 10]{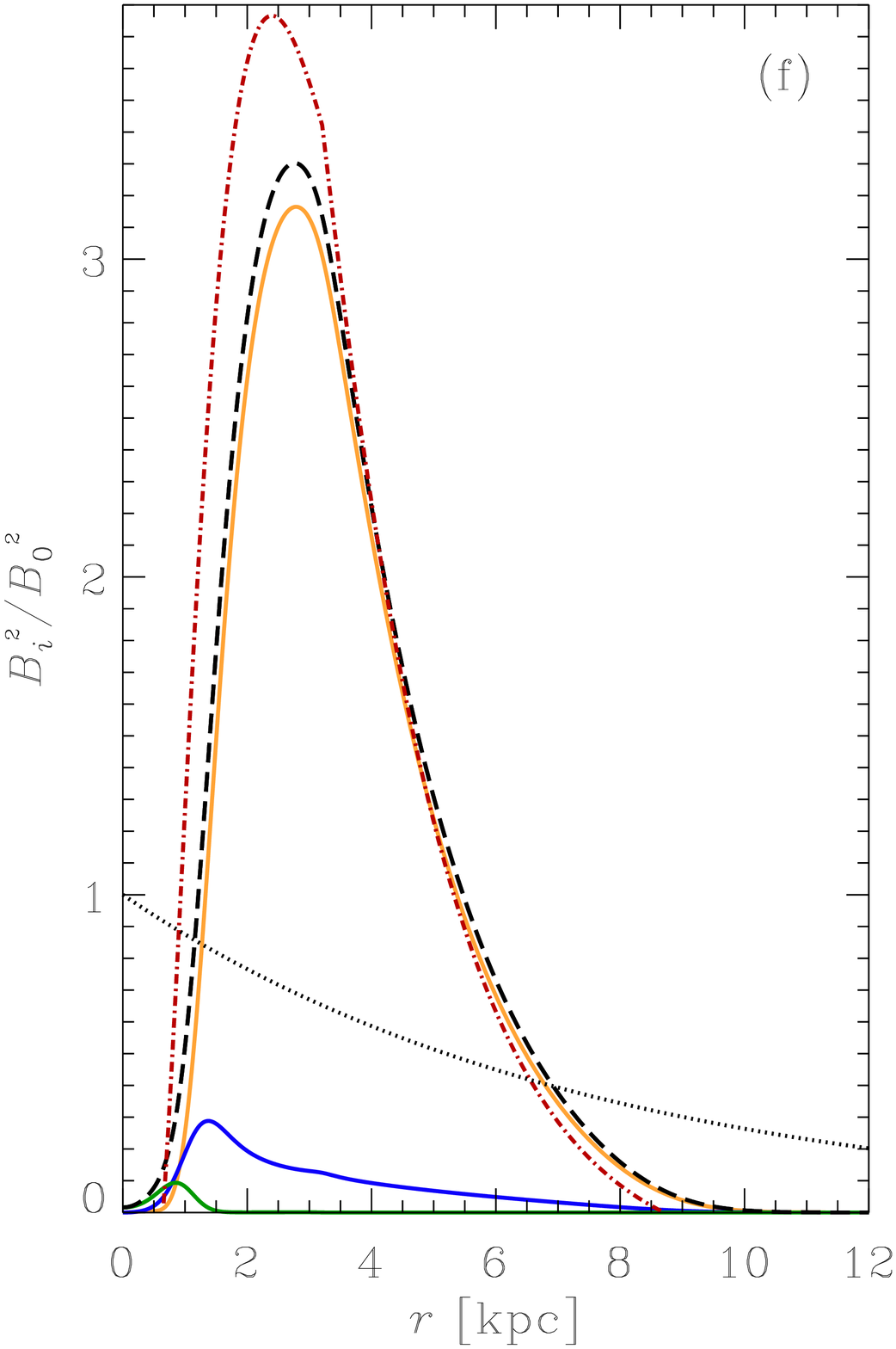}\\
  \includegraphics[width=58mm,clip=true,trim=  10 10 10 10]{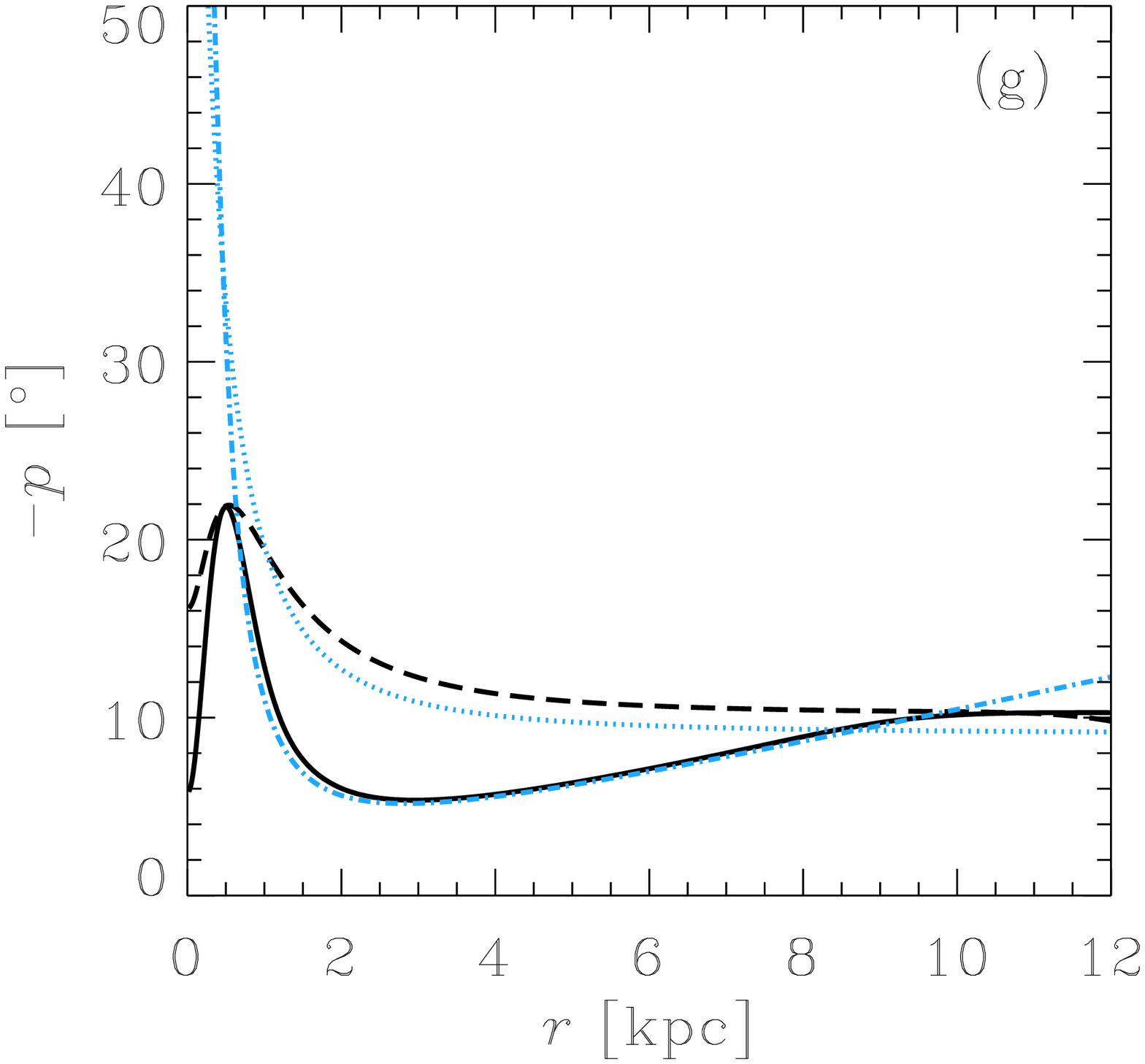}
  \includegraphics[width=58mm,clip=true,trim=  10 10 10 10]{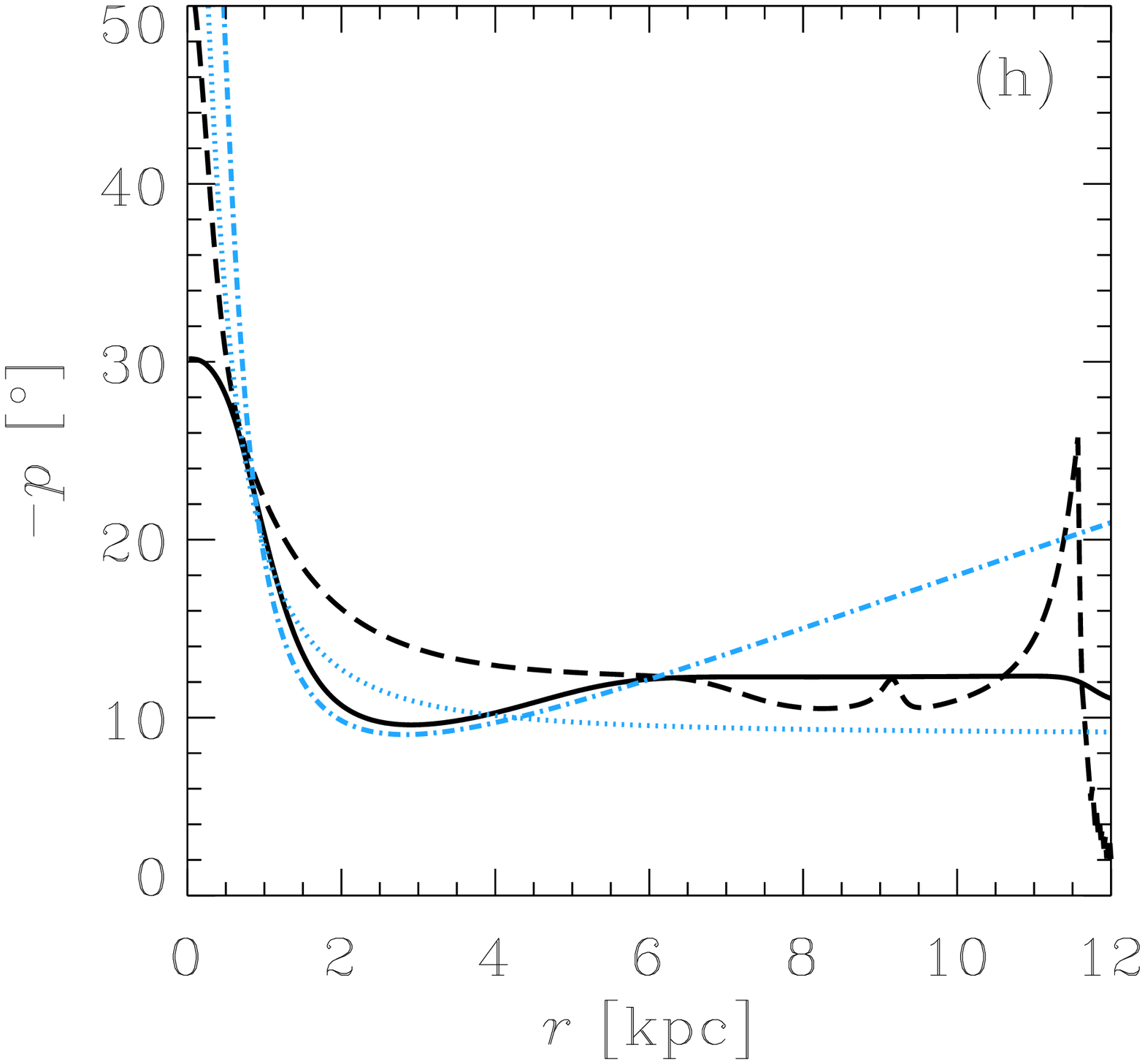}
  \includegraphics[width=58mm,clip=true,trim=  10 10 10 10]{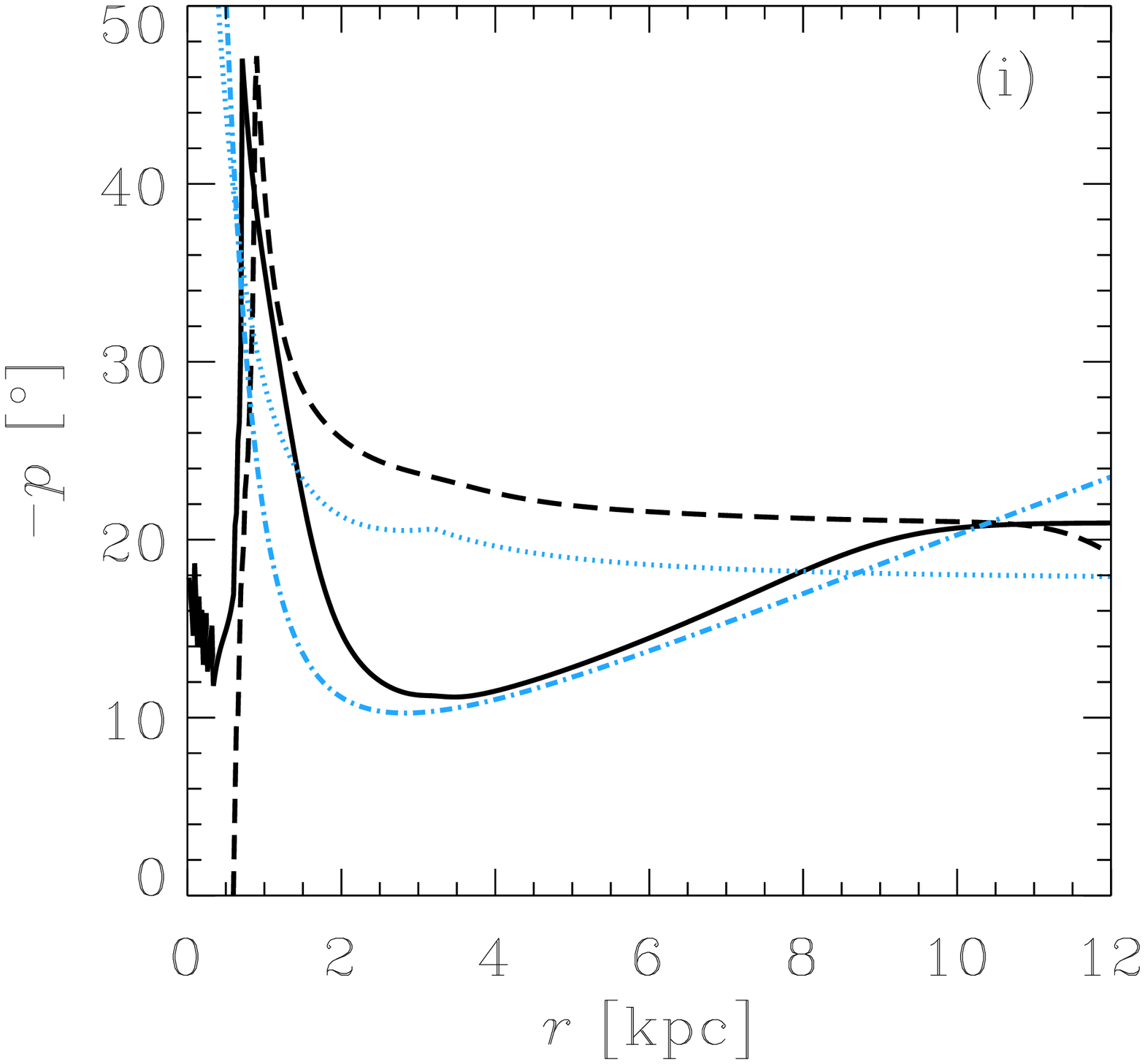}
  \caption{Solutions for the mean magnetic field for Model~A (left-most column), B (middle column), and C (right-most column).
           \textbf{Panels~(a)-(c):} Time evolution of the square-root of the normalized average magnetic energy density (solid).
                                    A line with slope equal to the value of the global growth rate $\Gamma$ 
                                    obtained using analytical theory is shown for reference (dashed).
           \textbf{Panels~(d)-(f)):} Normalized energy density in each component of $\meanv{B}$, averaged across the disc,
                                    in the saturated state ($t=15\Gyr$).
                                    The total energy density obtained 
                                    from the numerical solution $\langle\mean{B}^2\rangle/B\f^2$ (dashed), 
                                    is equal to the sum of the components, 
                                    $\langle\mbr^2\rangle/B\f^2$ (blue solid),
                                    $\langle\mbp^2\rangle/B\f^2$ (orange solid), and
                                    $\langle\mbz^2\rangle/B\f^2$ (green solid).
                                    The total can be compared with the analytical estimate~\eqref{Bsat}
                                    for $\langle\mean{B}^2\rangle/B\f^2$ (dashed-dotted).
                                    The midplane equipartition field $B\eq^2(r,0)/B\f^2$ (dotted) 
                                    is also shown for reference.
           \textbf{Panels~(g)-(i):} The pitch angle for the mean magnetic field.
                                    Average values across the disc $\langle p \mean{B}^2\rangle/\langle\mean{B}^2\rangle$ (black) 
                                    are shown for the kinematic regime ($t=1.5\Gyr$, $2\Gyr$, and $1\Gyr$, 
                                    respectively, for Models~A, B and C dashed) and saturated state ($t=15\Gyr$, solid).
                                    The analytical estimates for $p\kin$ (dotted) and $p\sat$ (dashed-dotted), 
                                    given by equations~\eqref{pkin} and \eqref{psat}, respectively, are also shown.
                                    Numerical artefacts in the solution for $p\kin$ in panel~(h) are inconsequential,
                                    as they occur outside the region where the field strength is significant.
                                    Sharp features in the numerically determined pitch angle in panel~(i)
                                    are caused by the definition of $p$ to lie between $-\pi/2$ and $\pi/2$,
                                    and are not physically meaningful.
           \label{fig:Bcompare}
          }            
\end{figure*}

\begin{figure}                     
  \includegraphics[width=87mm,clip=true,trim=  20 10 0 10]{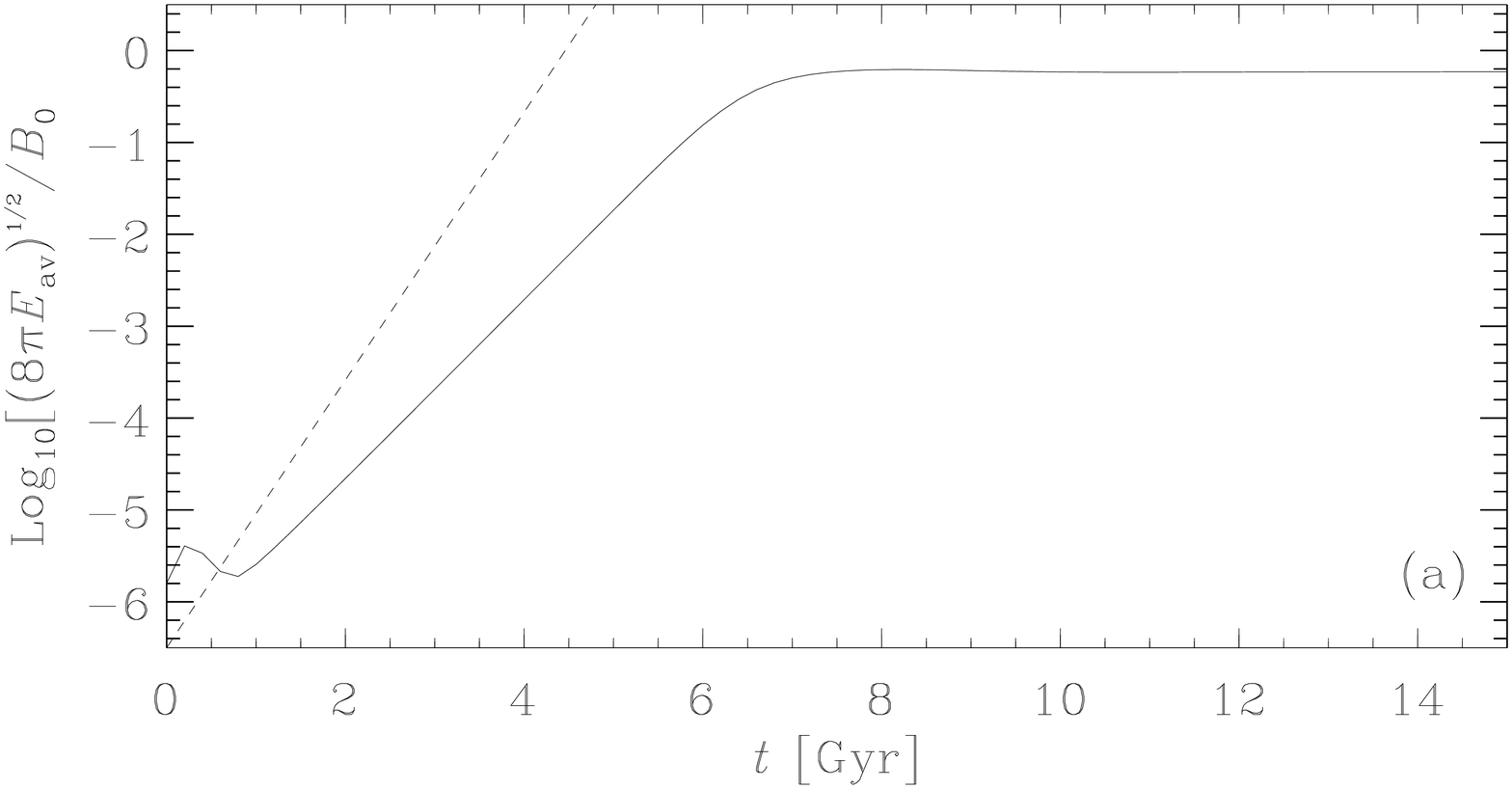}\\
  \includegraphics[width=87mm,clip=true,trim=  20 10 0 10]{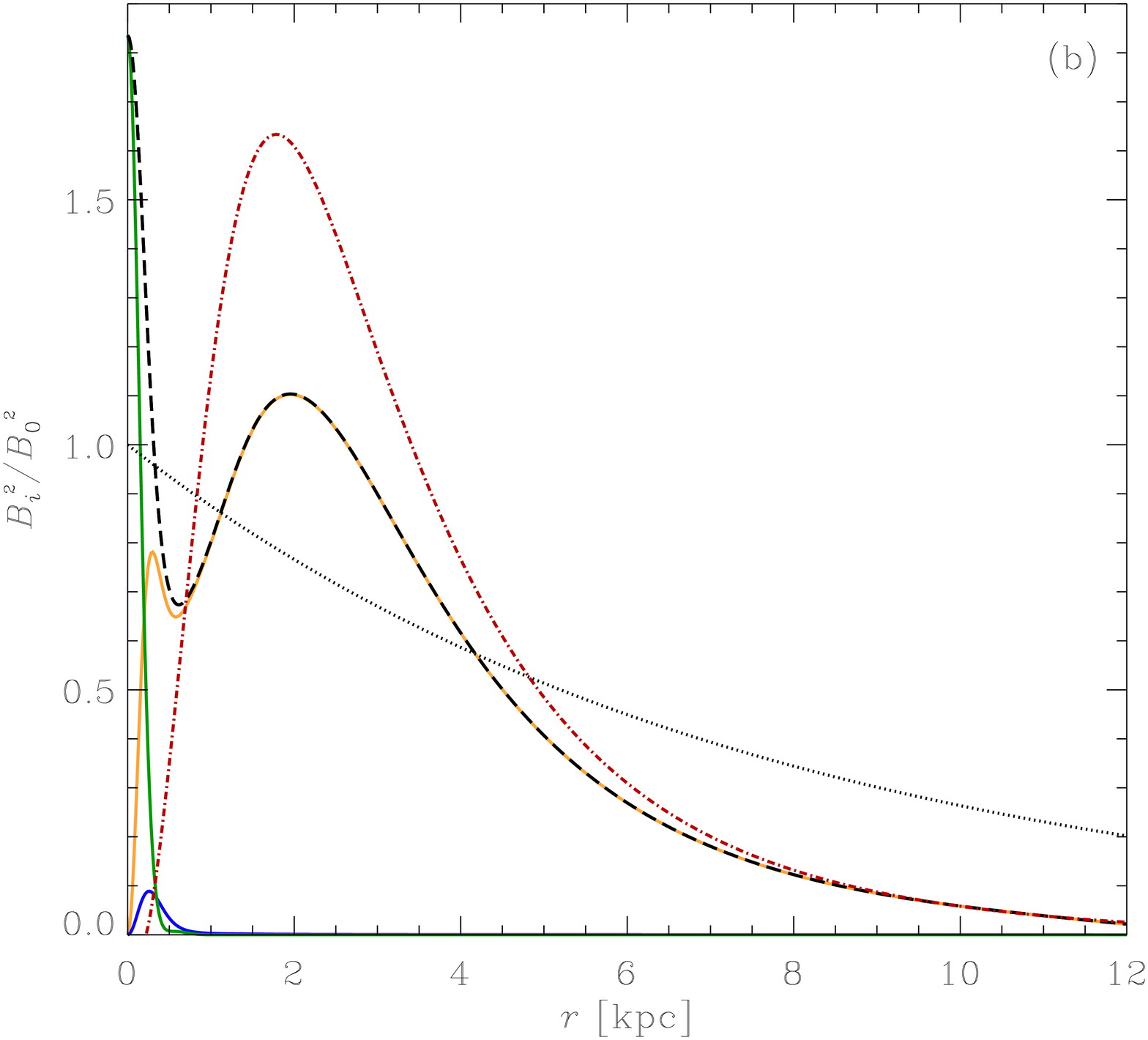}\\
  \includegraphics[width=87mm,clip=true,trim=  20 10 0 10]{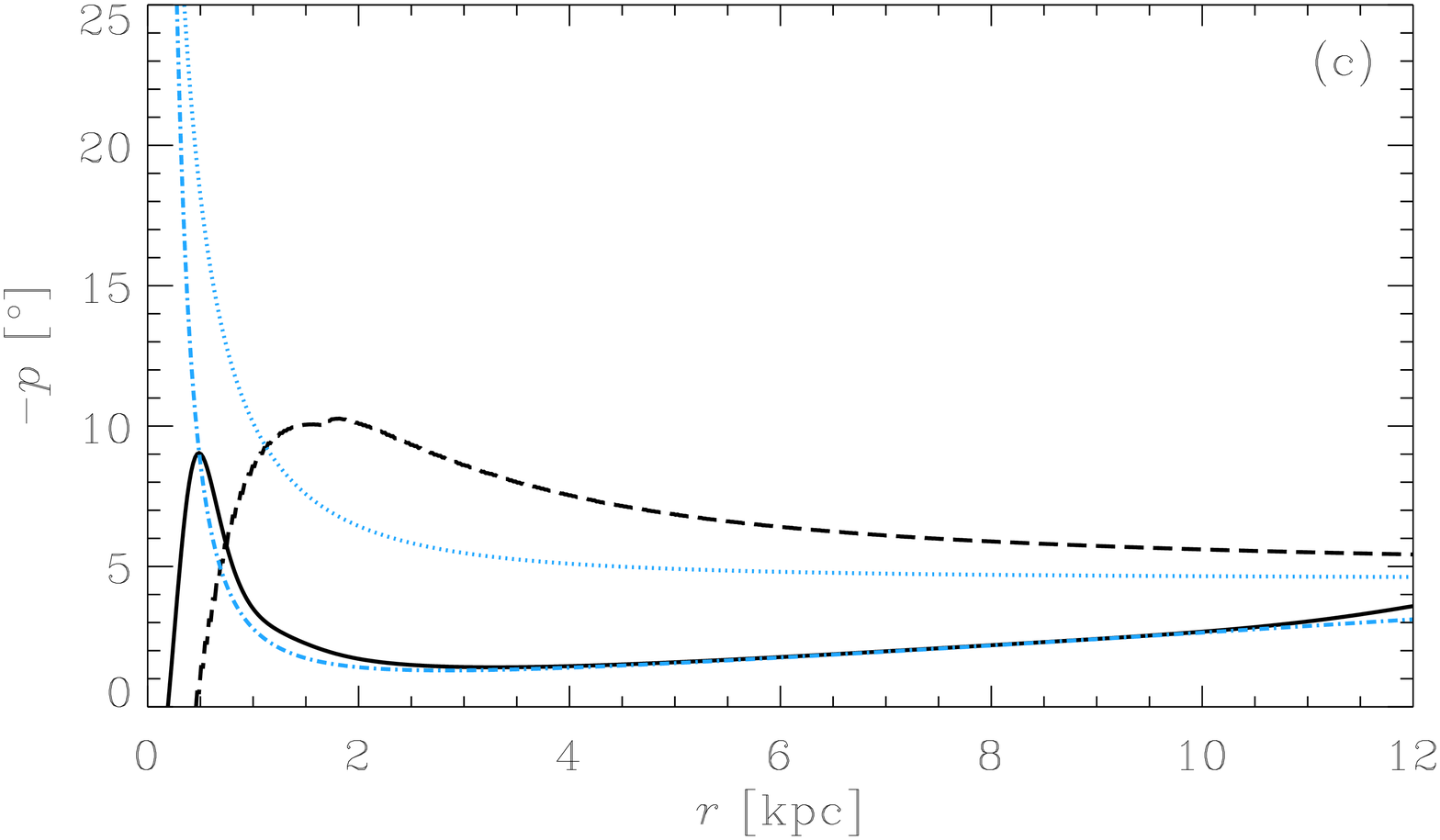}
  \caption{As Figure~\ref{fig:Bcompare}, but now for Model~D, which differs from Model~A in having a twice larger scale height $h$. 
           The time used to plot the kinematic regime is $t=3.4\Gyr$.
           \label{fig:Bcompare_D}
          }            
\end{figure}

\begin{figure*}
  \begin{minipage}[t][0mm][t]{\textwidth}
    \subfloat{\includegraphics[height=32.5mm,clip=true,trim=   32 -15 58 -30]{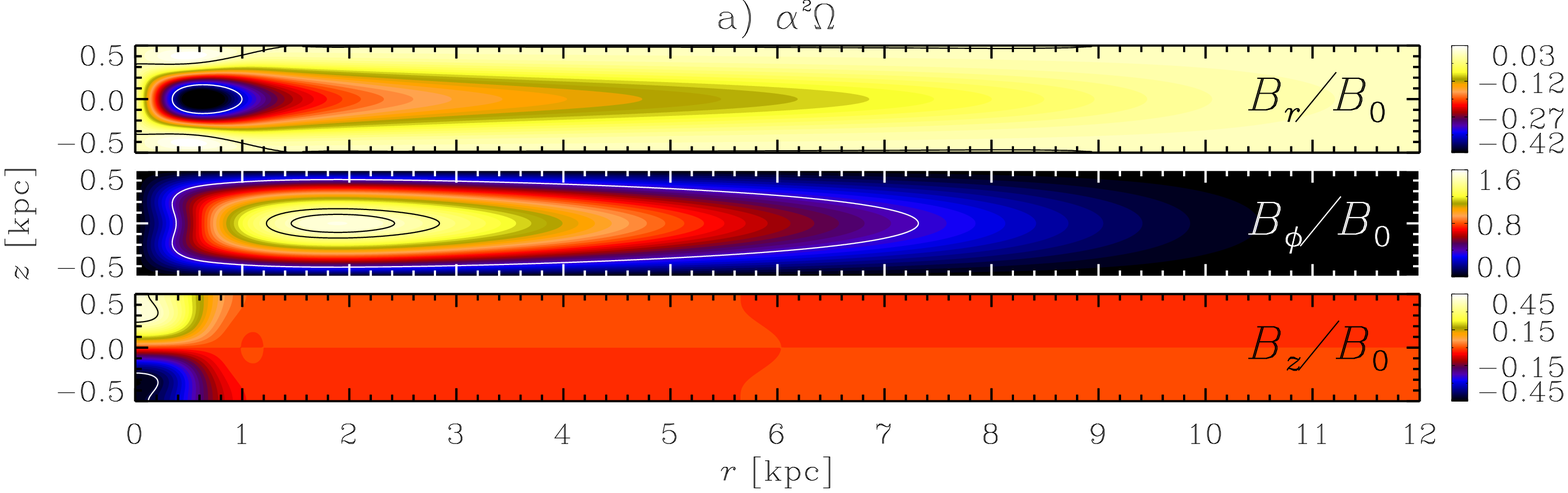}}
    \hspace{1mm}
    \subfloat{\includegraphics[height=32.5mm,clip=true,trim=   97 -15 -20 -30]{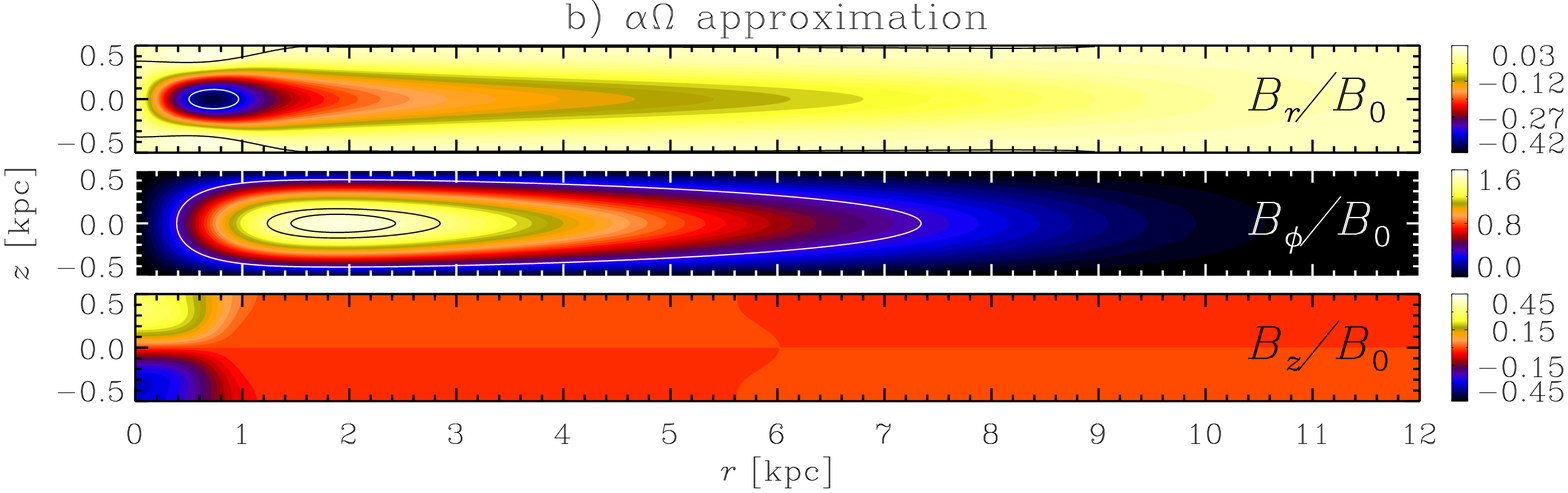}}\\
  \end{minipage}\\ \vspace{32mm}
  \begin{minipage}[t][0mm][t]{\textwidth}
    \subfloat{\includegraphics[height=32.5mm,clip=true,trim=   32 -15 58 -30]{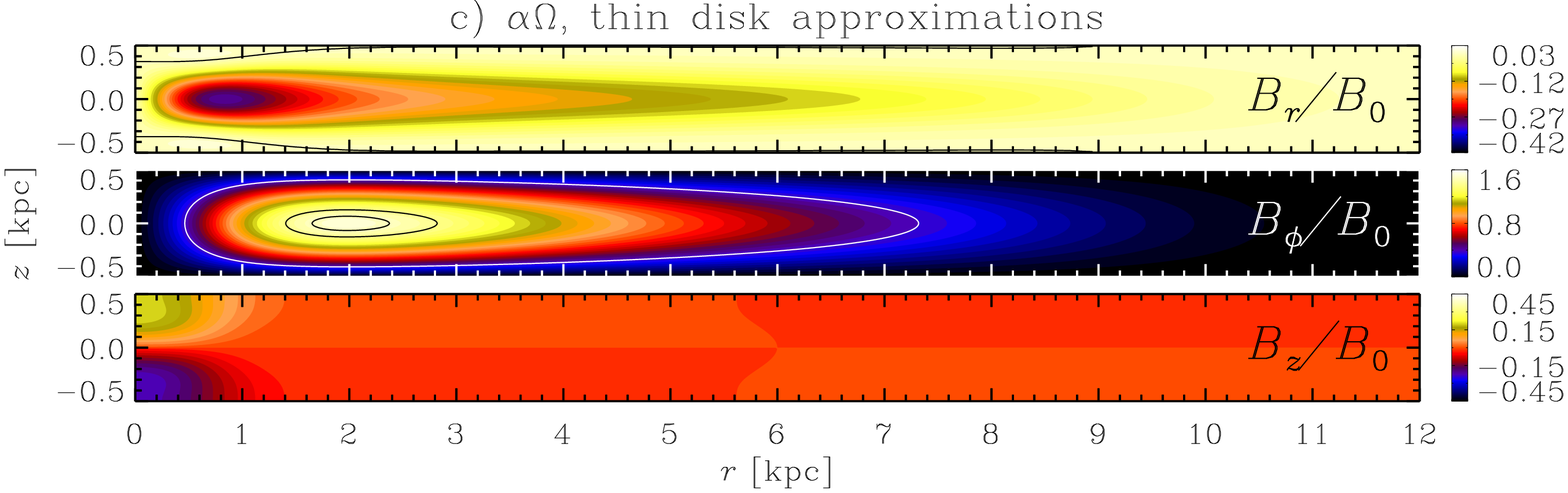}}	
    \hspace{1mm}
    \subfloat{\includegraphics[height=32.5mm,clip=true,trim=   97 -15 58 -30]{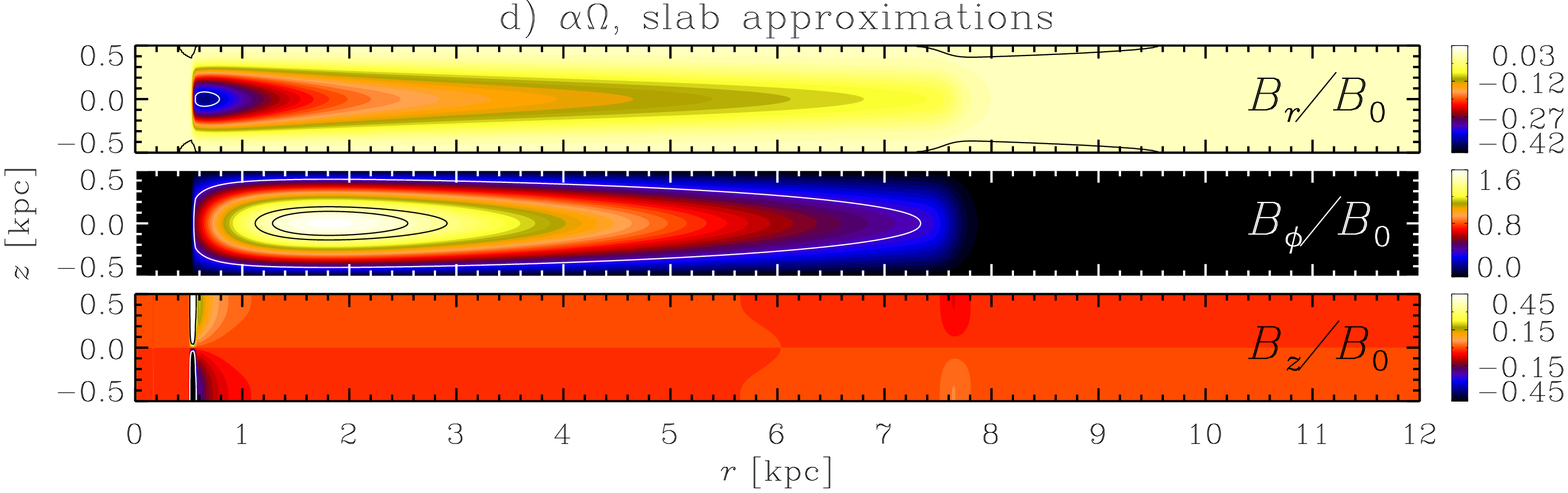}}\\
  \end{minipage}\\ \vspace{32mm}
  \begin{minipage}[t][0mm][t]{\textwidth}
    \subfloat{\includegraphics[height=32.5mm,clip=true,trim=   32 -15 58 -30]{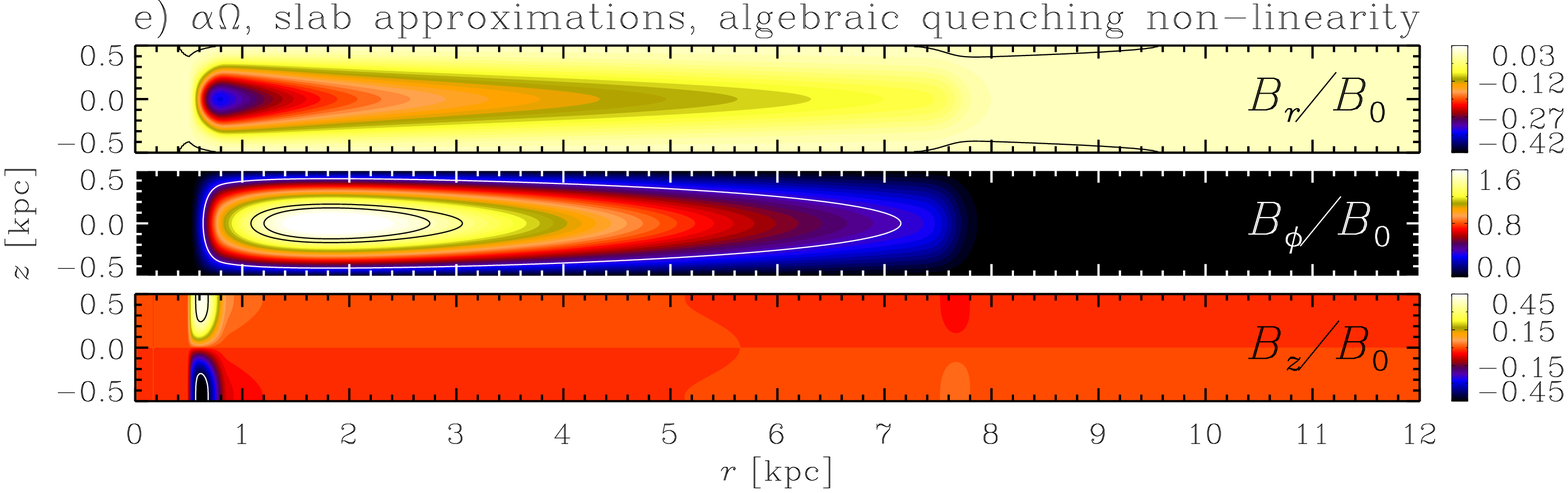}}
    \hspace{1mm}
    \subfloat{\includegraphics[height=32.5mm,clip=true,trim=   97 -15 58 -30]{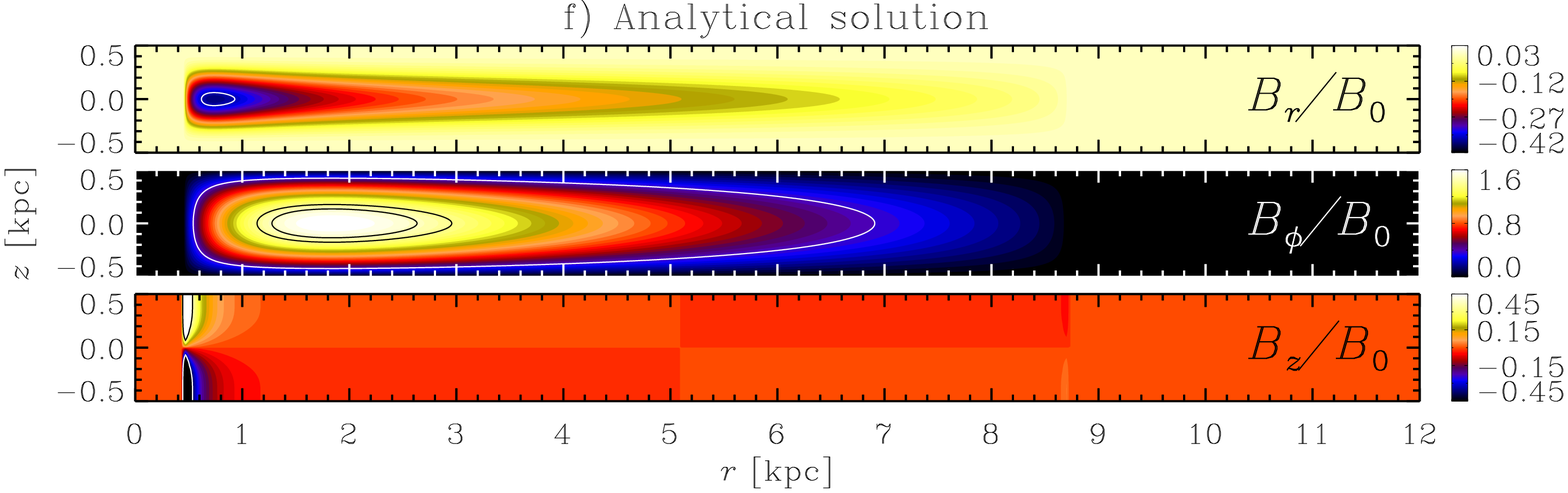}}
  \end{minipage}\\ \vspace{36mm}
  \caption{
    The value of each magnetic component $\mbr$, $\mbp$, and $\mbz$,
    normalized to the equipartition field strength at $(r,z)=(0,0)$, $B\f$, 
    shown as a function of the coordinates $r$ and $z$ (to scale),
    for the saturated (steady) state in Model~A.
    Colours were permitted to saturate in some cases.
    For clarity, contours have been drawn at $\mbr/B\f=-0.35$ and $0$, $\mbp/B\f=0.4$, $1.4$ and $1.5$, and $\mbz/B\f=-0.4$ and $0.4$.
    \textbf{Panel (a):} Full numerical solution.
    \textbf{Panel (b):} Numerical solution neglecting terms involving $\alpha$ in the equation for $\psi$ 
                        (the $\alpha$-$\Omega$ approximation).
    \textbf{Panel (c):} As panel (b) but now neglecting certain terms involving $\del\psi/\del r$,
                        that is, neglecting certain effects of $\mbz$ on $\mbr$ and $\mbp$,
                        and with $\mbz$ determined from $\mbr$ and $\bfDel\cdot\meanv{B}=0$ (the thin-disc approximation).
    \textbf{Panel (d):} As panel (c) but now neglecting all terms involving radial derivatives (the slab approximation).
    \textbf{Panel (e):} As panel (d) but now replacing the dynamical $\alpha$-quenching non-linearity 
                        with the generalized algebraic $\alpha$-quenching non-linearity (equation~\eqref{algebraic}).
    \textbf{Panel (f):} Combined perturbation and no-$z$ analytical solution 
                        described in Section~\ref{sec:combined}, 
                        using the same spatial resolution as the numerical solutions.
    \label{fig:Bicontour_A}
  }
\end{figure*}

\begin{figure*}
  \begin{minipage}[t][0mm][t]{\textwidth}
    \subfloat{\includegraphics[height=32.5mm,clip=true,trim=   32 -15 58 -30]{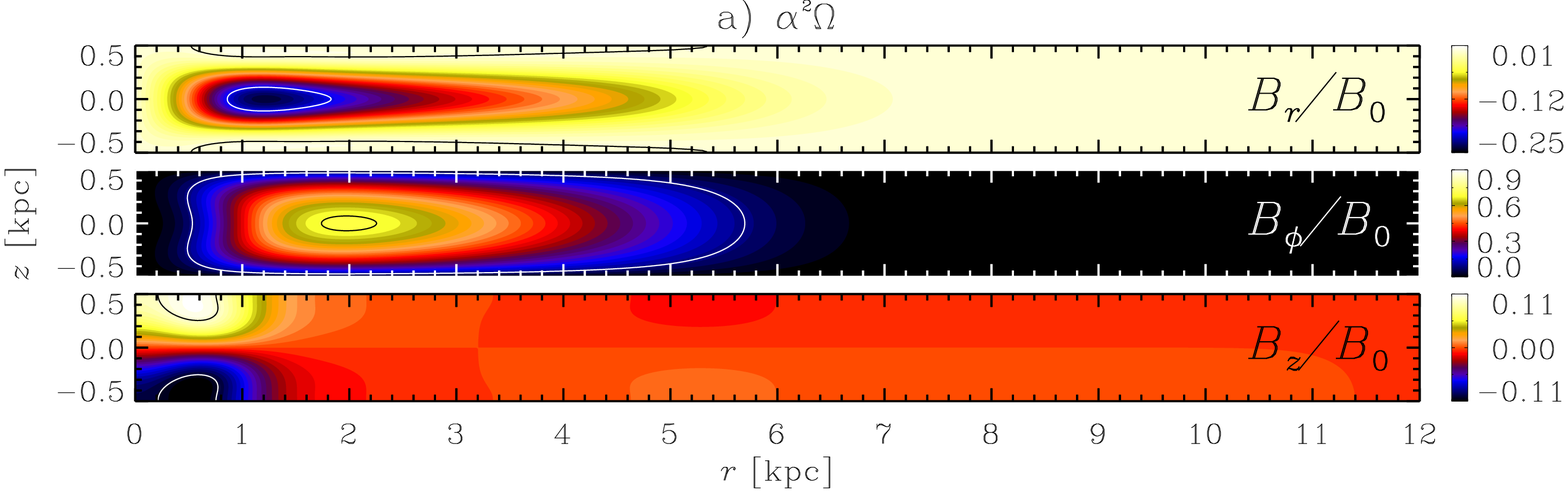}}
    \hspace{1mm}
    \subfloat{\includegraphics[height=32.5mm,clip=true,trim=   97 -15 -20 -30]{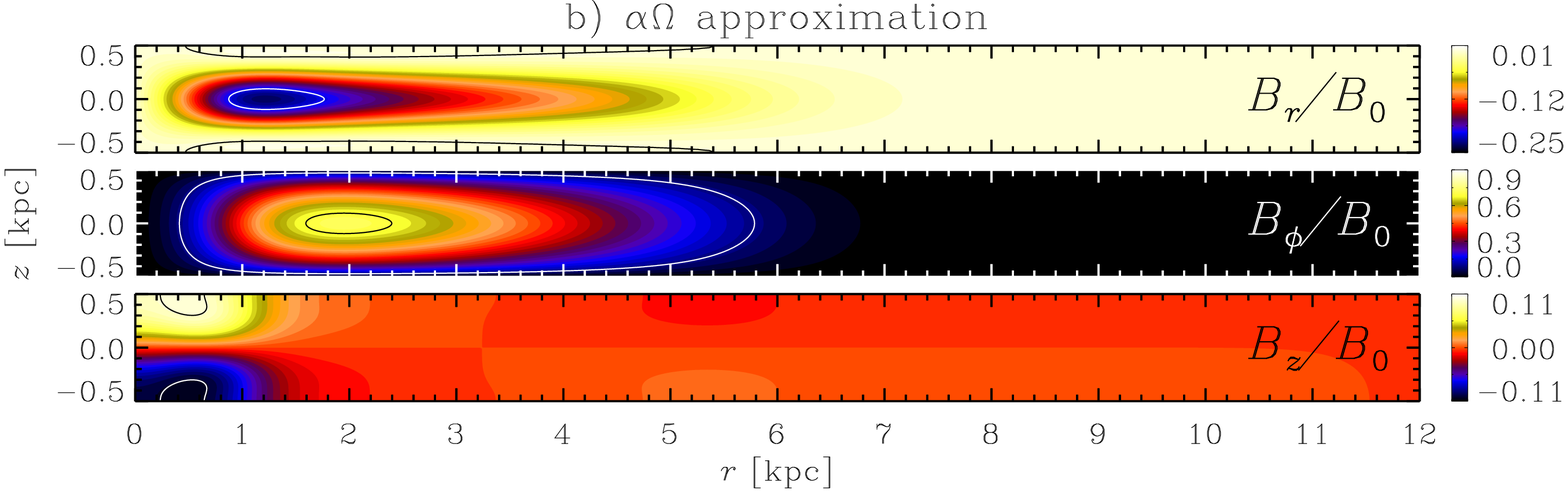}}
  \end{minipage}\\ \vspace{32mm}
  \begin{minipage}[t][0mm][t]{\textwidth}
    \subfloat{\includegraphics[height=32.5mm,clip=true,trim=   32 -15 58 -30]{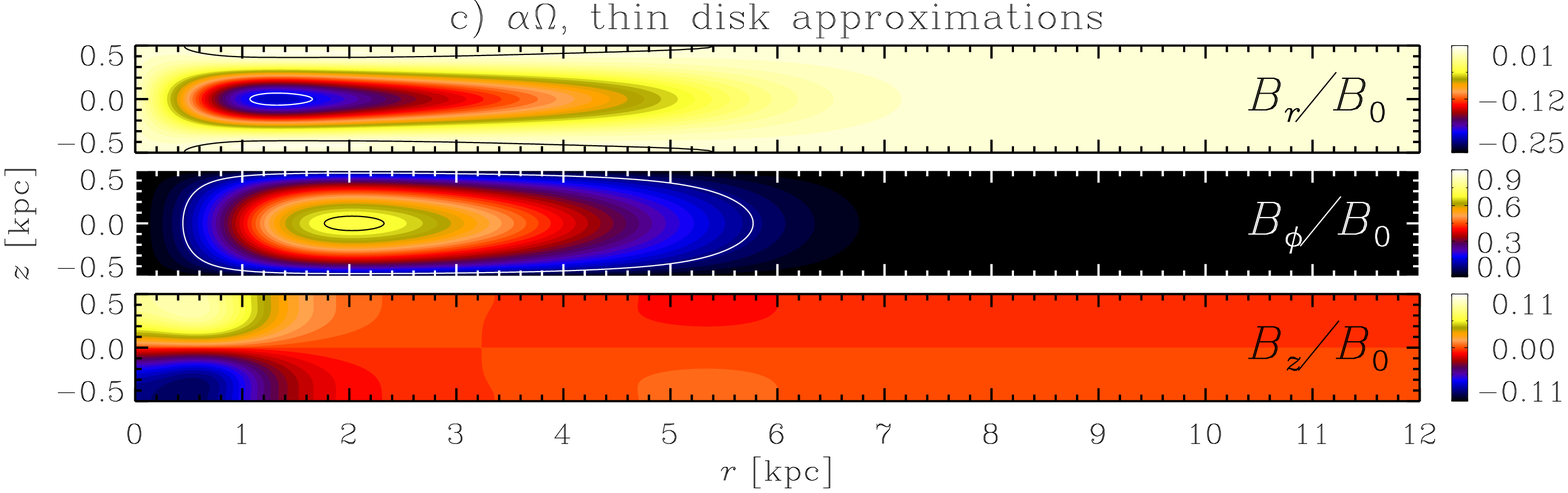}}
    \hspace{1mm}
    \subfloat{\includegraphics[height=32.5mm,clip=true,trim=   97 -15 58 -30]{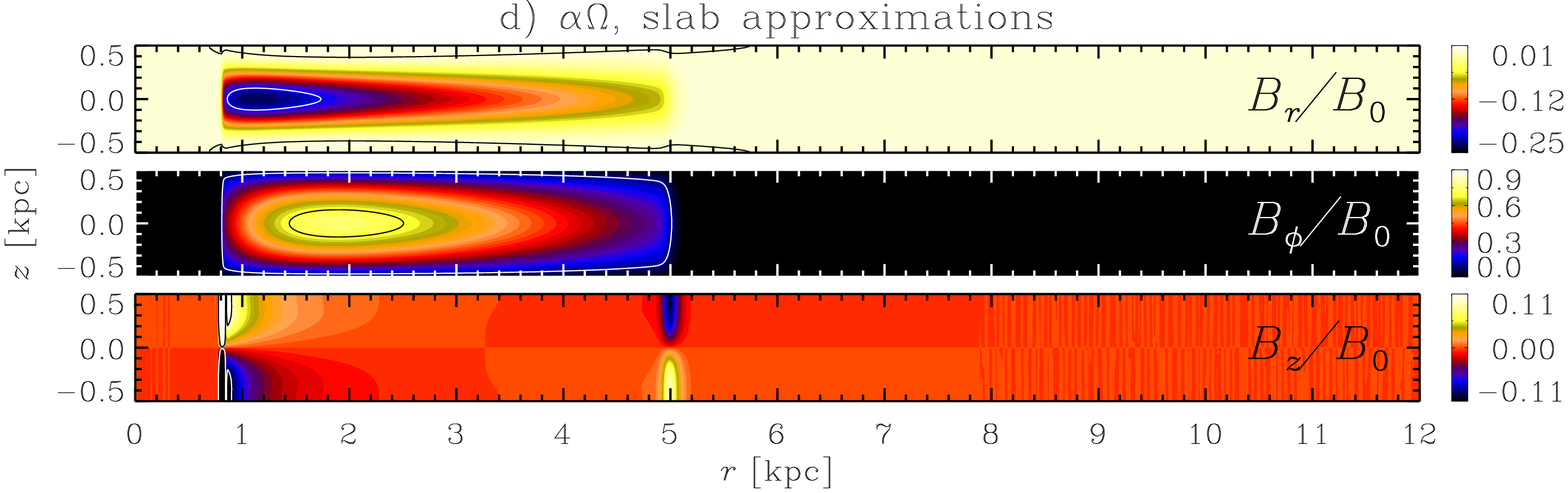}}
  \end{minipage}\\ \vspace{32mm}
  \begin{minipage}[t][0mm][t]{\textwidth}
    \subfloat{\includegraphics[height=32.5mm,clip=true,trim=   32 -15 58 -30]{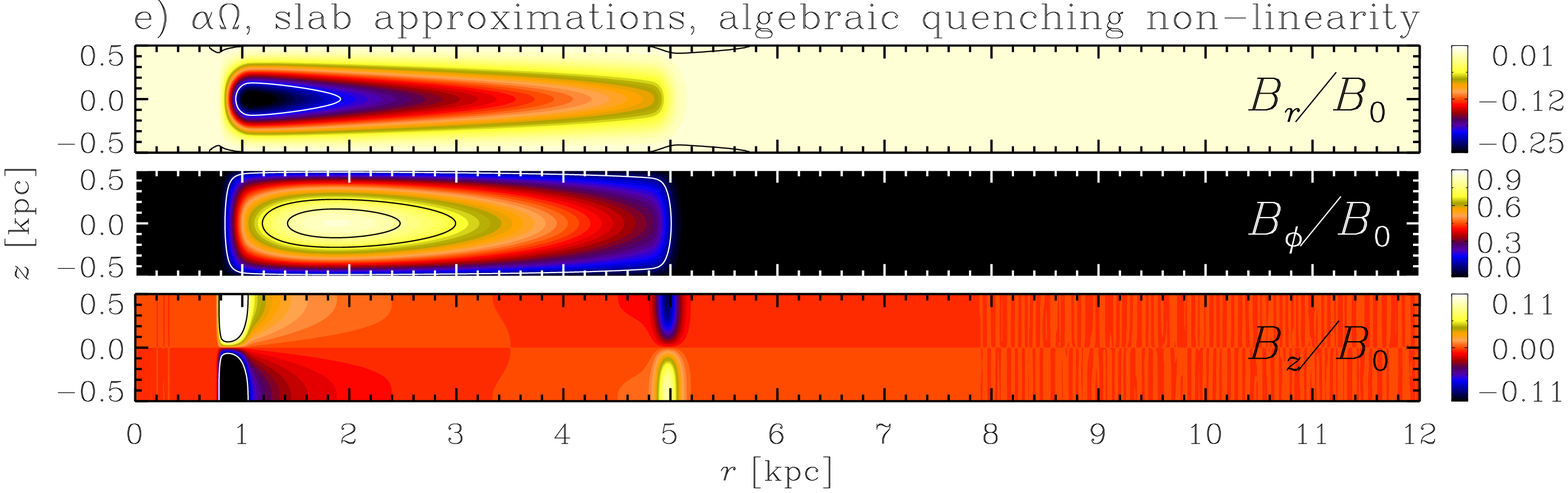}}
    \hspace{1mm}
    \subfloat{\includegraphics[height=32.5mm,clip=true,trim=   97 -15 58 -30]{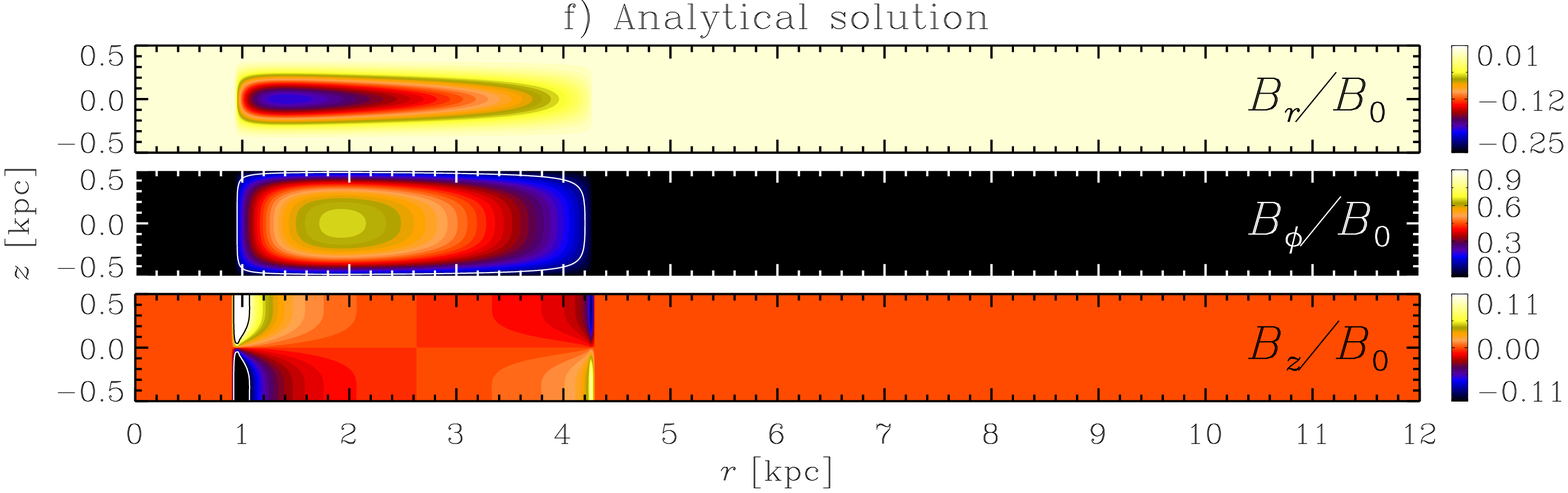}}
  \end{minipage}\\ \vspace{36mm}
  \caption{As Figure~\ref{fig:Bicontour_A} but now for Model~B, which has a strong outflow.
    For clarity, contours have been drawn at $\mbr/B\f=-0.2$ and $0$, $\mbp/B\f=0.1$, $0.7$ and $0.8$, and $\mbz/B\f=-0.1$ and $0.1$.
    \label{fig:Bicontour_B}
  }
\end{figure*}

\begin{figure*}
  \begin{minipage}[t][0mm][t]{\textwidth}
    \subfloat{\includegraphics[height=32.5mm,clip=true,trim=   32 -15 58 -30]{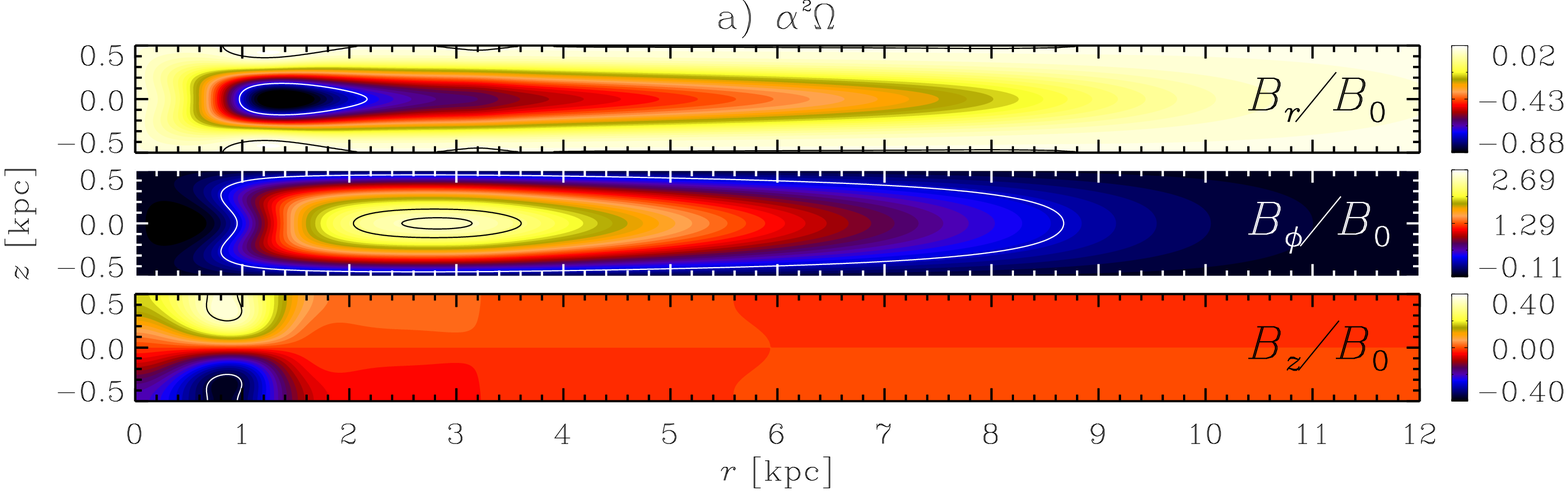}}
    \hspace{1mm}
    \subfloat{\includegraphics[height=32.5mm,clip=true,trim=   97 -15 -20 -30]{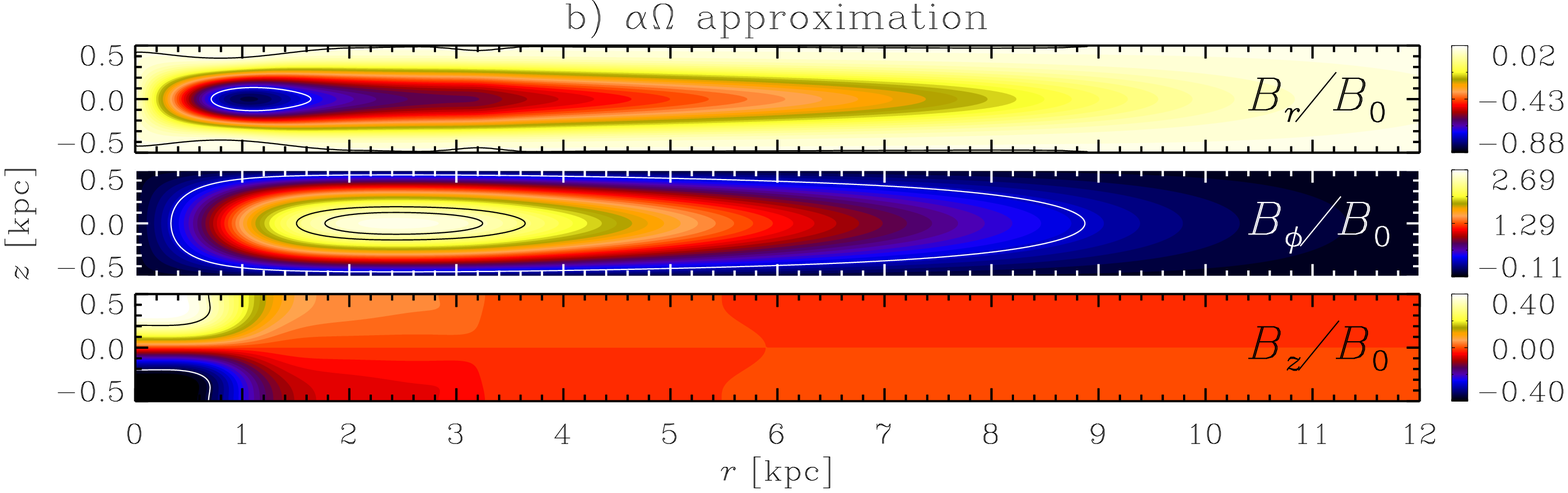}}
  \end{minipage}\\ \vspace{32mm}
  \begin{minipage}[t][0mm][t]{\textwidth}
    \subfloat{\includegraphics[height=32.5mm,clip=true,trim=   32 -15 58 -30]{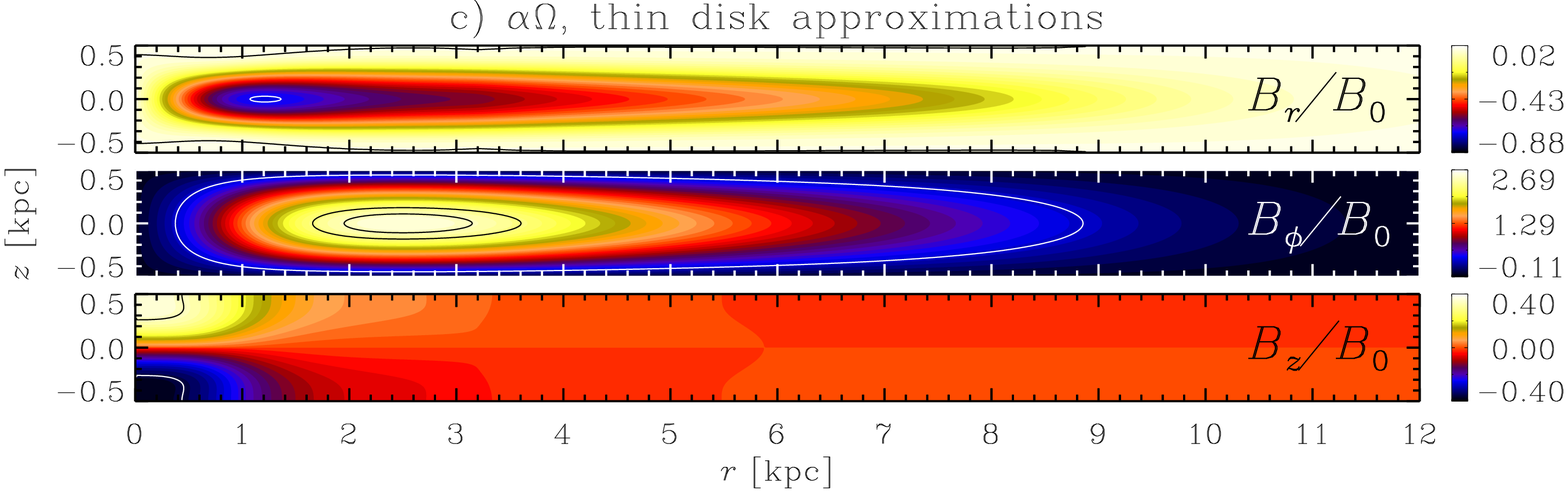}}
    \hspace{1mm}
    \subfloat{\includegraphics[height=32.5mm,clip=true,trim=   97 -15 58 -30]{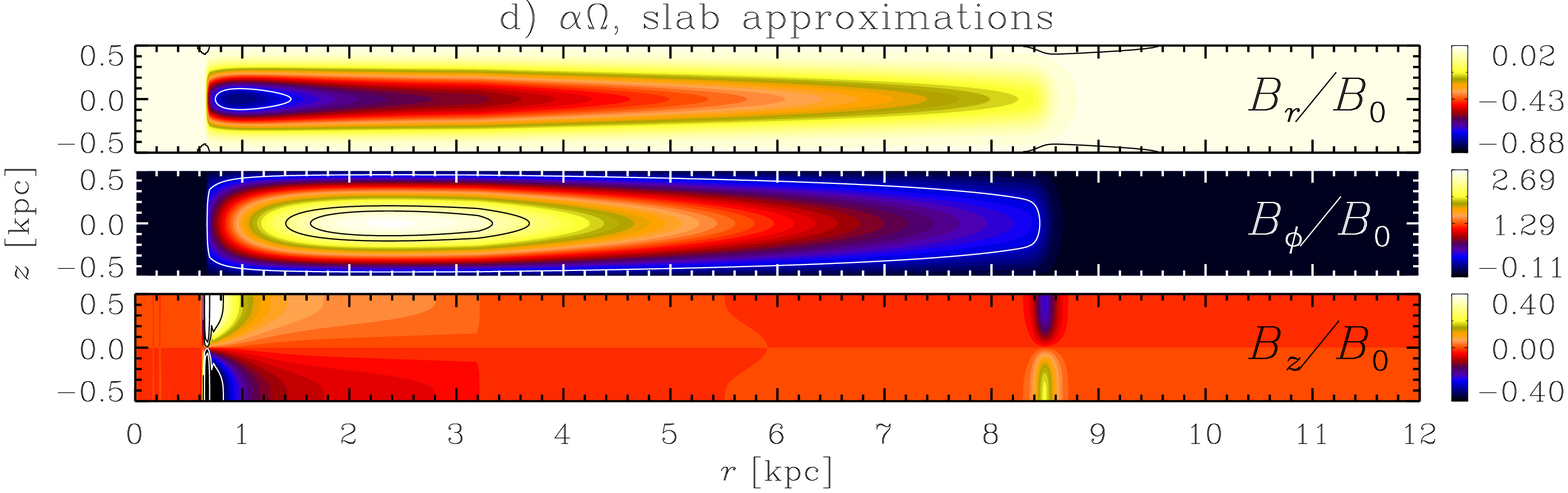}}
  \end{minipage}\\ \vspace{32mm}
  \begin{minipage}[t][0mm][t]{\textwidth}
    \subfloat{\includegraphics[height=32.5mm,clip=true,trim=   32 -15 58 -30]{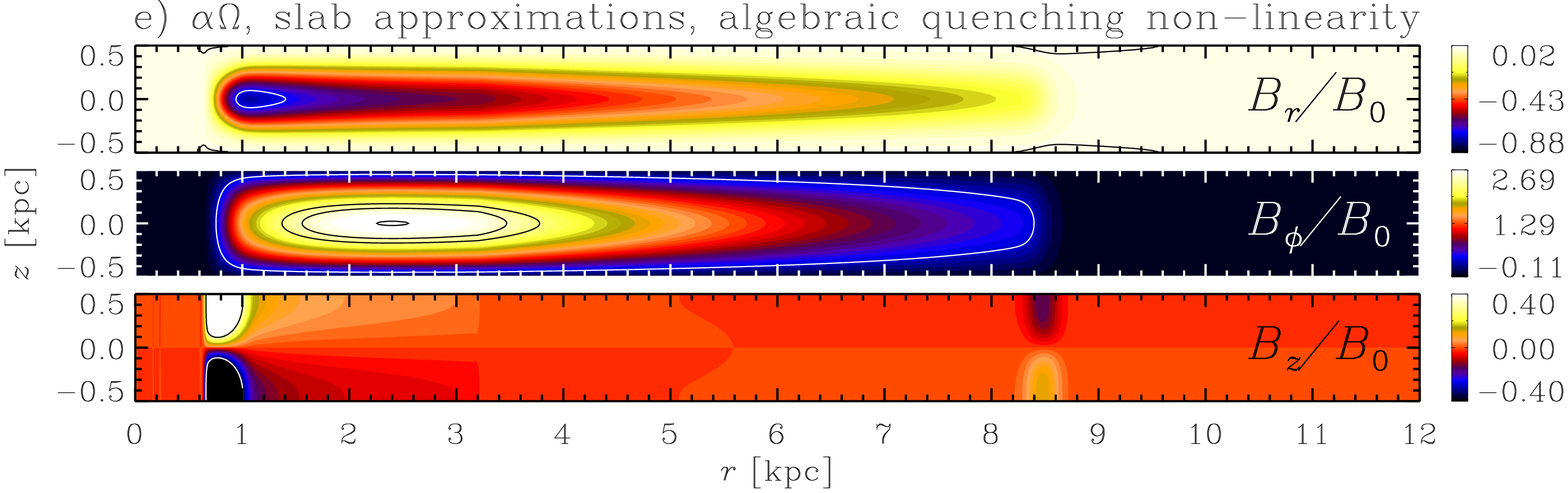}}
    \hspace{1mm}
    \subfloat{\includegraphics[height=32.5mm,clip=true,trim=   97 -15 58 -30]{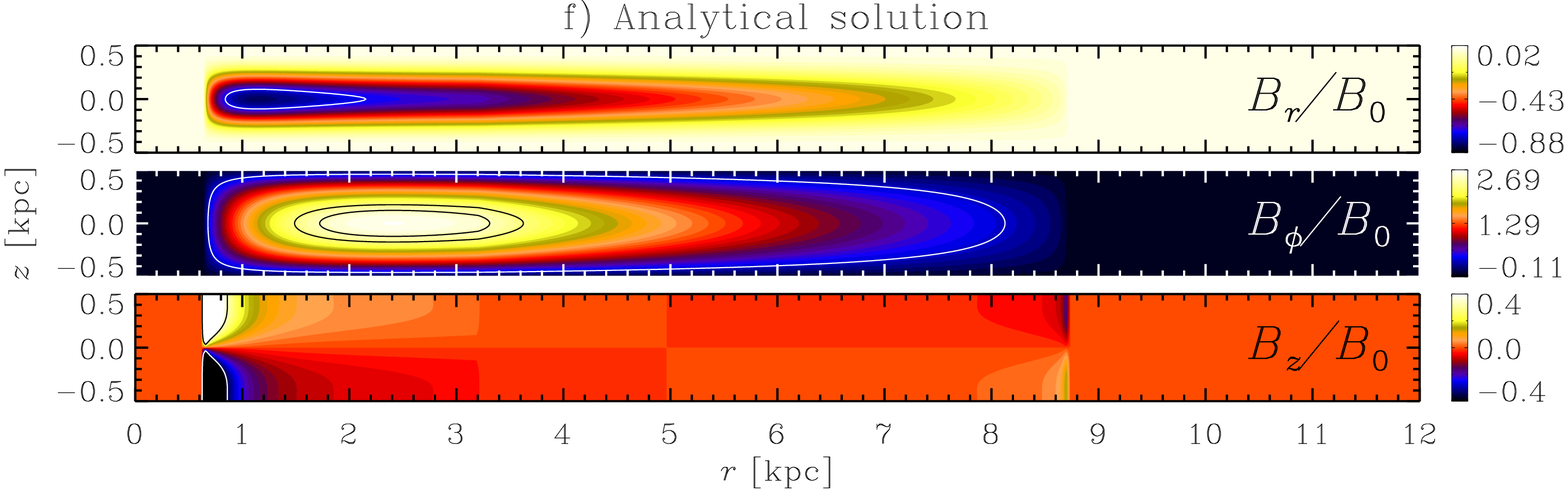}}
  \end{minipage}\\ \vspace{36mm}
  \caption{As Figure~\ref{fig:Bicontour_A} but now for Model~C, which has a larger turbulent scale $l=\tau u$.
           For clarity, contours have been drawn at $\mbr/B\f=-0.7$ and $0$, 
           $\mbp/B\f=0.35$, $2.3$, $2.5$ and $2.8$, and $\mbz/B\f=-0.35$ and $0.35$.
    \label{fig:Bicontour_C}
  }
\end{figure*}

\begin{figure*}
  \begin{minipage}[t][0mm][t]{\textwidth}
    \subfloat{\includegraphics[height=60mm,clip=true,trim=   32 -15 58 -30]{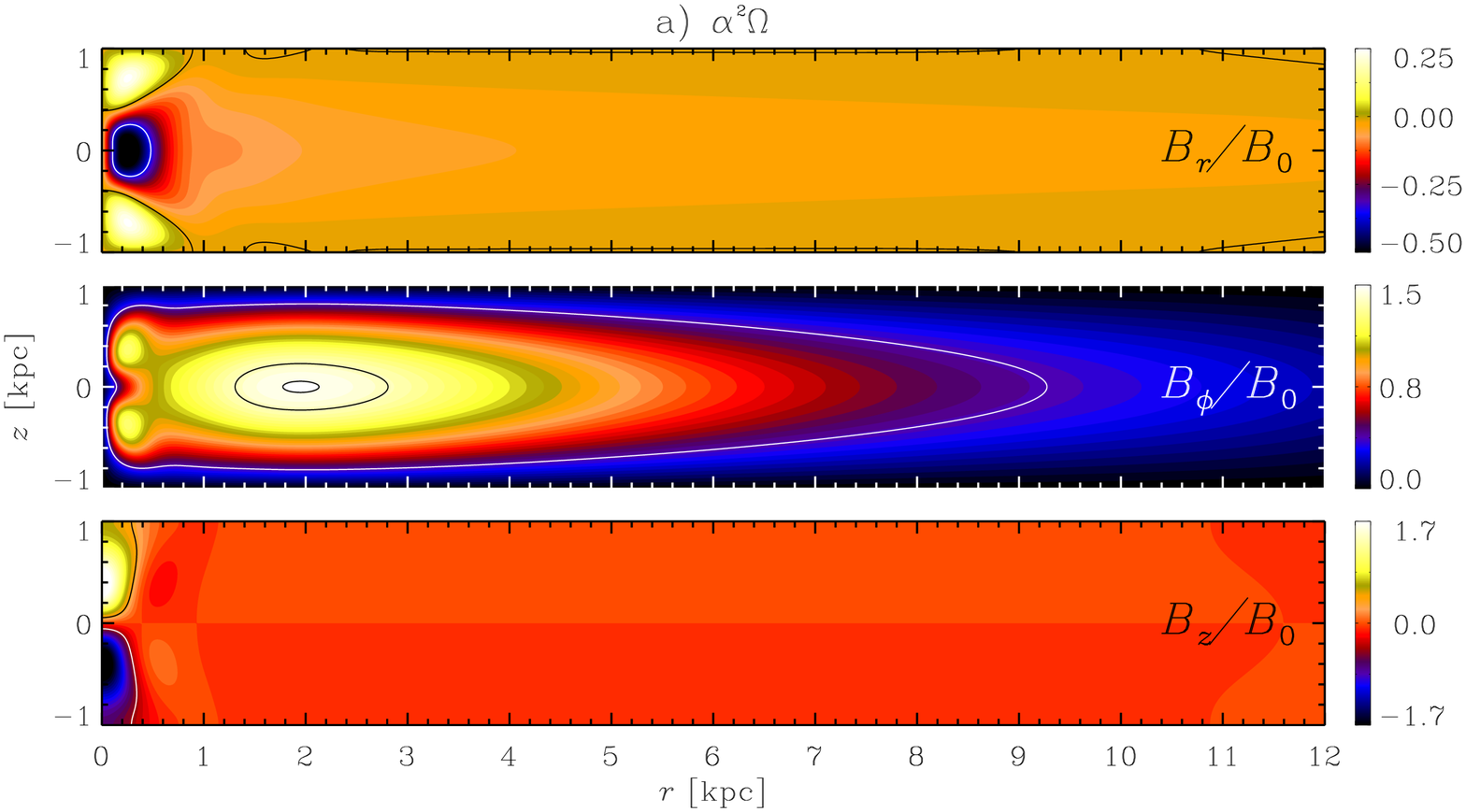}}
    \hspace{2mm}
    \subfloat{\includegraphics[height=60mm,clip=true,trim=   97 -15 -20 -30]{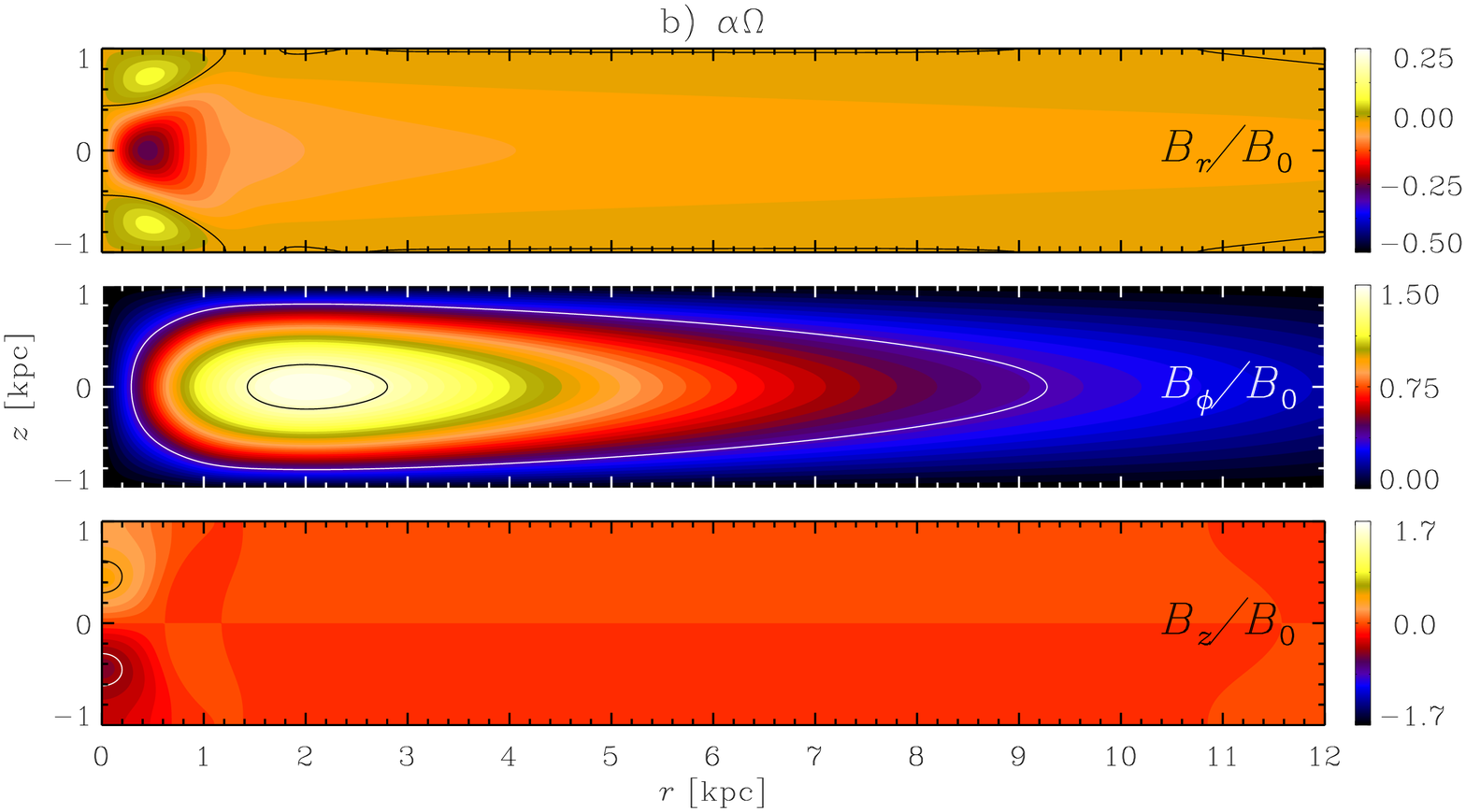}}
  \end{minipage}\\ \vspace{56mm}
  \begin{minipage}[t][0mm][t]{\textwidth}
    \subfloat{\includegraphics[height=60mm,clip=true,trim=   32 -15 58 -30]{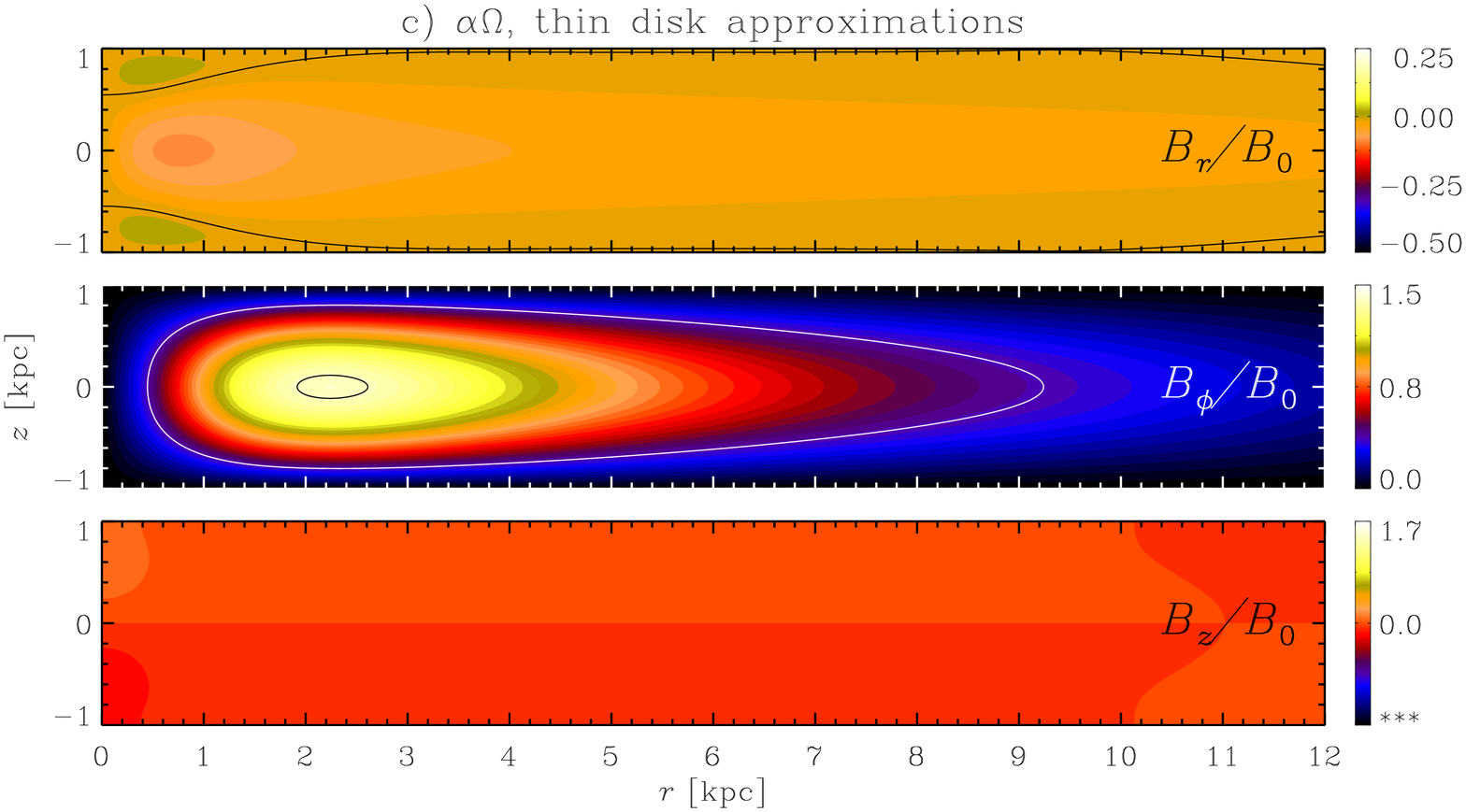}}
    \hspace{2mm}
    \subfloat{\includegraphics[height=60mm,clip=true,trim=   97 -15 58 -30]{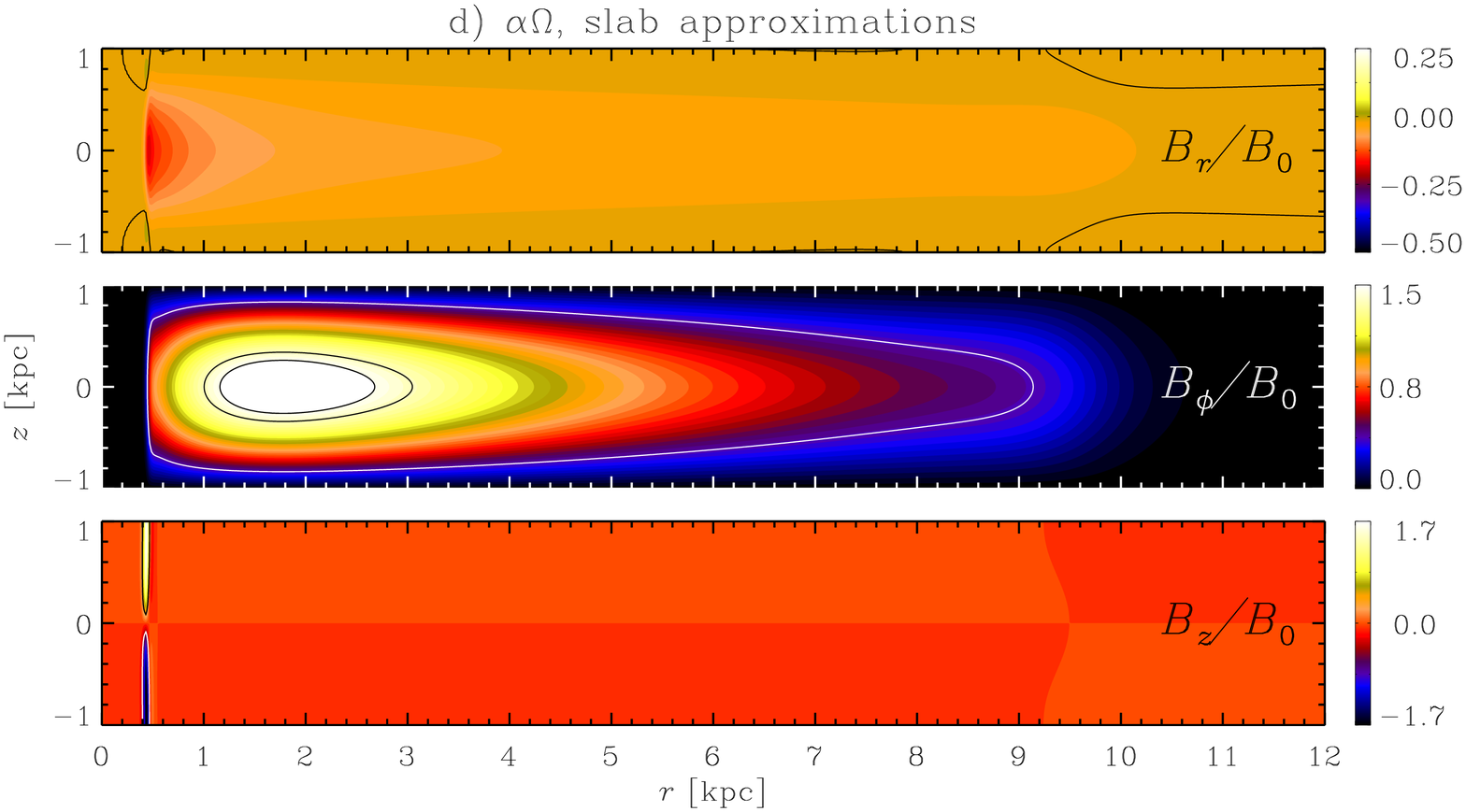}}
  \end{minipage}\\ \vspace{56mm}
  \begin{minipage}[t][0mm][t]{\textwidth}
    \subfloat{\includegraphics[height=60mm,clip=true,trim=   32 -15 58 -30]{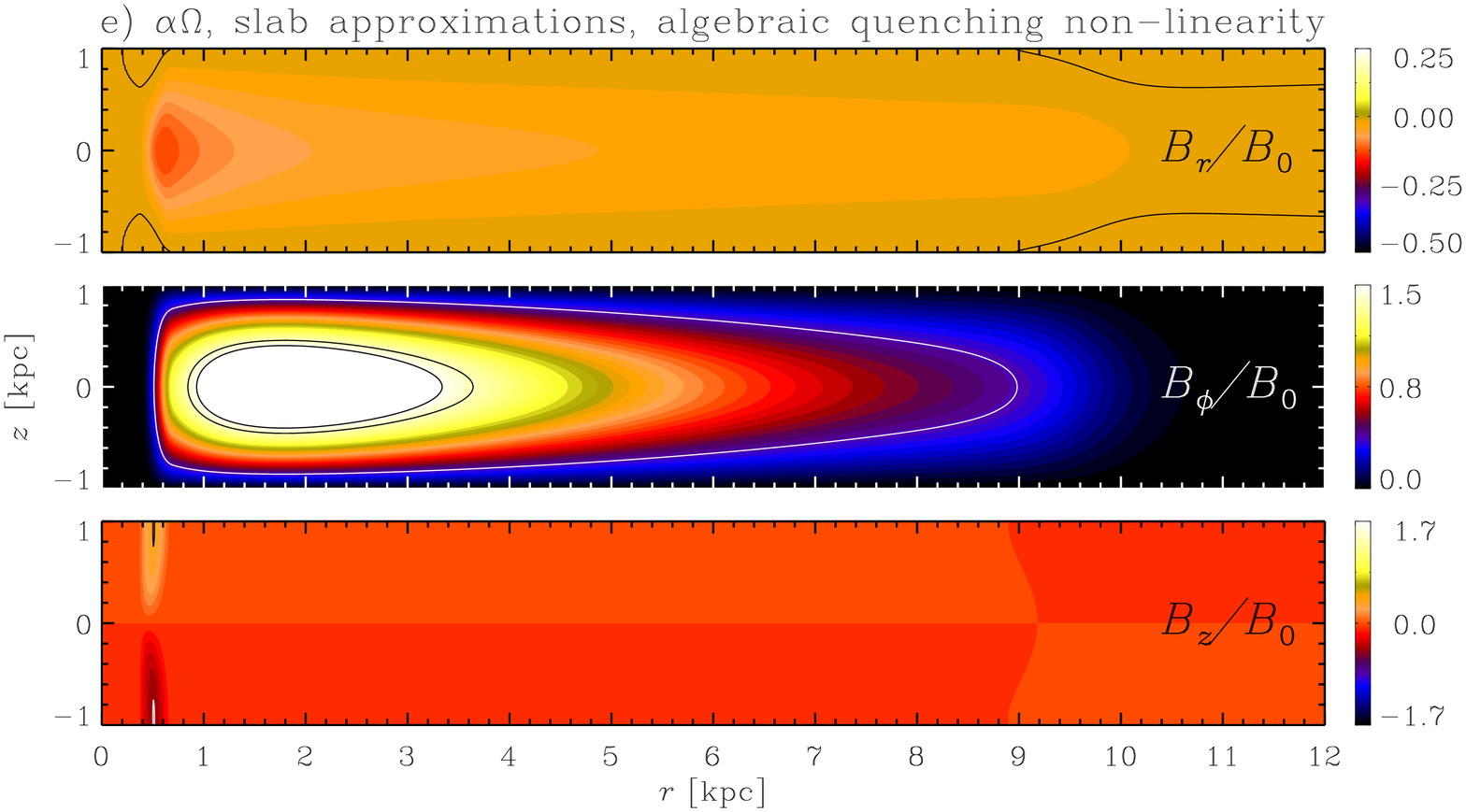}}
    \hspace{2mm}
    \subfloat{\includegraphics[height=60mm,clip=true,trim=   97 -15 58 -30]{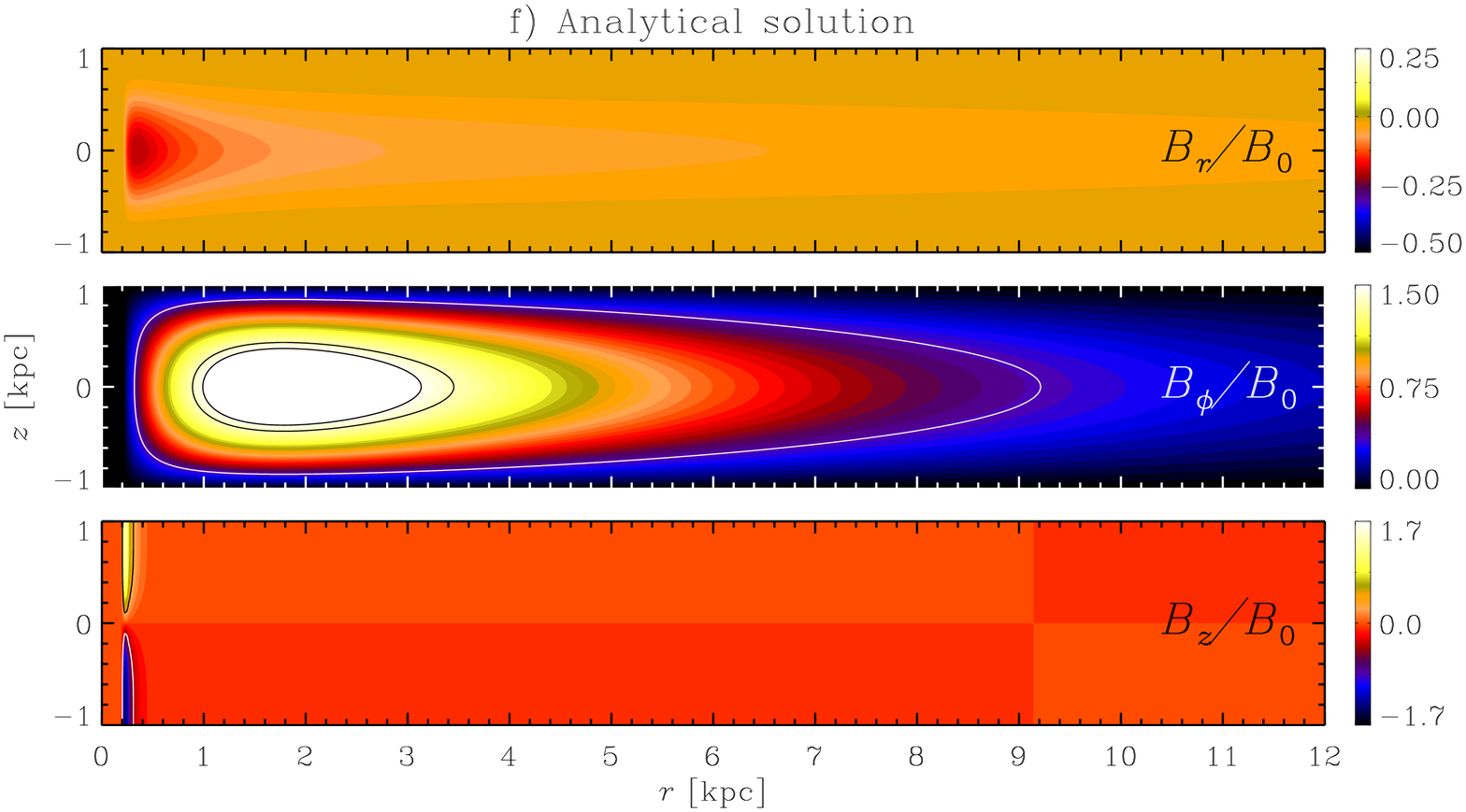}}
  \end{minipage}\\ \vspace{62mm}
  \caption{As Figure~\ref{fig:Bicontour_A} but now for Model~D, which has a larger scale height $h$.
    Contours have been drawn at $\mbr/B\f=-0.35$ and $0$, $\mbp/B\f=0.4$, $1.4$ and $1.5$, 
    and $\mbz/B\f=-0.4$ and $0.4$, as drawn for Model~A in Figure~\ref{fig:Bicontour_A}.
    \label{fig:Bicontour_D}
  }
\end{figure*}

\section{Results}
\label{sec:results}
We now present full numerical solutions for the galactic disc dynamo,
and compare these with numerical solutions obtained under various approximations,
as well as with the analytical solution discussed in Section~\ref{sec:analytic}.
We consider four cases, labelled as Models~A to D,
which have the parameters listed in Table~\ref{tab:models}.
The fiducial model, Model~A, has no outflow and has standard parameters $h=0.5\kpc$, $l=0.1\kpc$ and $u=12\kms$.
Model~B differs from Model~A only in that it has a fairly strong outflow, with $U\f=3\kms$
(for comparison, the field decays if $U\f\gtrsim6\kms$).
Model~C, on the other hand, has the same parameters as the fiducial model except that $l=0.2\kpc$.
This causes $\Coriolis>1$ for $r<3.2\kpc$, 
which means that $\alpha\kin$ is rotationally saturated (equation \eqref{alpha_0}) in Model~C for $r<3.2\kpc$.
For other models, $\alpha\kin$ decreases monotonically with $r$.
It reaches a maximum amplitude of $2.3\kms$ at $r=0$ for Models~A and B,
and $1.1\kms$ for Model~D,
while for Model~C, $\alpha\kin$ has a constant amplitude of $4.8\kms$ between $r=0$ and $r=3.2\kpc$.
Model~D is like Model~A except that $h$ is doubled form $0.5\kpc$ to $1\kpc$.

Disc dynamo solutions are shown in Figures~\ref{fig:Bcompare}--\ref{fig:Bicontour_D}.
Figure~\ref{fig:Bcompare} shows data that is vertically averaged across the disc, 
with each column of panels corresponding to a given model, 
from left to right: A, B, and C,
while Figure~\ref{fig:Bcompare_D} shows the same information for Model~D.
Figures~\ref{fig:Bicontour_A}--\ref{fig:Bicontour_D}
show the full spatial dependence of the mean field in the steady state for Models~A--D.
Axes are shown to scale to emphasize the aspect ratios,
and the $r$-axis is truncated for presentational convenience.
To begin with we consider three quantities: the kinematic growth rate of the mean magnetic field,
and the energy and pitch angle of the mean magnetic field in the saturated state.

\subsection{Kinematic growth rates}
\label{sec:gamma}
The time evolution of the square-root of the mean magnetic energy density over the simulation domain 
is plotted in panels~(a), (b), and (c) of Figure~\ref{fig:Bcompare} 
and panel~(a) of Figure~\ref{fig:Bcompare_D}.
This quantity is normalized to the equipartition value at the origin $B\f\equiv B\eq(0,0)$.
The magnetic energy initially decreases with time for the seed field adopted, 
but after a few hundred $\!\Myr$ the fastest growing eigenmode begins to dominate and the field then grows exponentially.
The final solution is not sensitive to the seed field chosen as long as it is sufficiently weak compared to equipartition.
The magnetic energy saturates $\sim2$--$7\Gyr$ after the start of the simulation, 
after growing $\sim10$--$12$ orders of magnitude, depending on the model.
The global growth rate $\Gamma$ in the kinematic regime is $4.8\Gyr^{-1}$ for Model~A, 
$3.6\Gyr^{-1}$ for Model~B, $8.2\Gyr^{-1}$ for Model~C, 
and $2.2\Gyr^{-1}$ for Model~D.
This corresponds to `$1000$-folding' times of $1.4$, $1.9$, $0.8$, 
and $3.1\Gyr$, respectively.

The analytical estimate for $\Gamma$, obtained using equation~\eqref{eigen} and shown by a dashed line, 
is $4.5\Gyr^{-1}$ for Model~A, which is remarkably close to the actual value.
For Model~B, however, the analytical estimate is $1.4\Gyr^{-1}$, or about $2/5$ of the actual value.
The underestimate for the large $V$ case is not surprising,
and can be traced to inaccuracies introduced by making the no-$z$ approximation 
on terms containing $\muz$ \citepalias{Chamandy+14b,Chamandy+Taylor15}.
In particular, the no-$z$ solution cannot be simultaneously calibrated to give accurate pitch angles 
\textit{and} accurate growth rates for strong outflows.
For Model~C, the analytical estimate is $7.5\Gyr^{-1}$, which is also remarkably close to the actual value,
while for Model~D the WKBJ treatment yields $3.4\Gyr^{-1}$, or $50\%$ larger than the actual value.
Analytical solutions rely on approximating the disc as thin, 
so are expected to be less accurate for Model~D, which has a thicker disc.

\subsection{Saturated field strengths}
\label{sec:strength}
In panels~\ref{fig:Bcompare}(d)--(f) and \ref{fig:Bcompare_D}(b),
the dashed lines show the normalized magnetic energy density averaged across the disc $\langle \mean{B}^2\rangle/B\f^2$,
plotted against radius, in the saturated regime ($t=15\Gyr$).
The normalized equipartition field $B\eq^2(r,0)/B\f^2$ is shown as a dotted line for reference.
The field is of approximately equipartition strength, and a few times larger for Model~A than for Model~B,
as the stronger outflow in Model~B leads to a less efficient dynamo.
This, in turn, is explained by referring back to equations~\eqref{dTdt} and \eqref{dpsidt}. 
Written in terms of $\mbi$, where $i$ represents $r$ or $\phi$,
and concerning ourselves only with the first terms on the right hand side,
we have
\begin{equation}
  \frac{\del\mbi}{\del t}=\ldots -\frac{\del\muz}{\del z}\mbi -\muz\frac{\del\mbi}{\del z}.
\end{equation}
The first term on the right-hand side is the expansion term, and leads to a reduction of $\mbi$ in our model,
whereas the second term is the advection term and leads to an enhancement of $\mbi$.
Though both terms are important, the expansion term turns out to be larger, leading to an overall reduction of $\mbi$.
For the same reason, the saturated magnetic energy tapers off more rapidly with radius for Model~B than for Model~A.
The magnetic field of Model~C, meanwhile, saturates at a strength a few times larger than that of Model~A.
Note that $\mbp^2$ (orange) dominates over $\mbr^2$ (blue) and $\mbz^2$ (green) in all models, as expected, 
except at $r\lesssim1\kpc$, where $\mbz^2$ may exceed $\mbp^2$ and $\mbr^2$
(see also Figures~\ref{fig:Bicontour_A}--\ref{fig:Bicontour_D}).
For Model~D, the vertically averaged magnetic energy actually peaks at $r=0$,
due to the strong vertical field there,
but $\mbz$ is much weaker for other models.

The analytical solution~\eqref{Bsat} for $\mean{B}\sat^2$ is shown as a dashed-dotted line 
in panels~\ref{fig:Bcompare}(d)--(f) and \ref{fig:Bcompare_D}(b), 
and generally gives a remarkably accurate estimate of the saturation field strength.
In Model~C, the agreement is slightly worse for $r$ smaller than the radius at which the field strength peaks.
This is caused by the neglect of the $\alpha^2$ effect in the analytical model;
when this effect is excluded from the mean-field simulation, the agreement is much better.
The $\alpha^2$ effect is more important for larger $l$ or $\tau=l/u$,
which was also found in \citetalias{Chamandy+Taylor15}.
Therefore, the increased importance of the $\alpha^2$ effect with $l$ 
seems to persist even when rotational saturation of $\alpha\kin$ is implemented.
Note also that the analytical solution, being local, does not taken into account the radial transport.
Therefore, it cannot be applied at radii $r<r_1$ or $r>r_2$ 
where $r_1$ and $r_2$ are the radii for which the normalized dynamo number $\Dtilde=1$, 
and where the analytical field strength goes to zero in each plot.
Moreover, the domain for which $\Dtilde>1$ is generally underestimated by the analytical solution.
This is because the solution overestimates the damaging effect of the outflow in the kinematic regime
(as mentioned above when discussing $\Gamma$), 
and also because the $\alpha^2$ effect and terms involving $\mbz$ are ignored in the analytical solution.

We now turn to the two-dimensional plots
of Figures~\ref{fig:Bicontour_A}--\ref{fig:Bicontour_D},
which show the field components for the saturated state, under various approximations, described in Section~\ref{sec:approx}.
We begin by comparing the full $\alpha^2\Omega$ numerical solutions of panels~(a) for Models~A--D.
The obvious difference in going from Model~A to the strong outflow model~B, 
is that the mean magnetic field becomes confined to a smaller radial domain,
because of the damaging effects of the outflow.
The component $\mbp$ is still peaked at about $r=2\kpc$, where the dynamo number peaks,
but the components $\mbr$ and $\mbz$, whose generation depends on the $\alpha$-effect, 
are peaked at somewhat larger radii compared to Model~A because the outflow weakens the dynamo at small $r$.
In Model~C, the peak of the dynamo number occurs farther out at about $r=2.8\kpc$ due to the rotational saturation of $\alpha\kin$,
and the peak of $\mbp$ is located at about the same radius.
The components $\mbr$ and $\mbz$ are also pushed out to larger radius compared to Models~A and B.
These features can also be seen in the middle row of Figure~\ref{fig:Bcompare}.
We have also run a variation of Model~C without rotational saturation,
but do not show the figures for the sake of brevity.
As expected, the field in that case resembles qualitatively that of Model~A at small radius,
e.g. in that its components respectively peak at roughly the same radii as in Model~A.
However, the profile of $\mbp$ retains the `double-tail' morphology at small radius (dip at the midplane)
seen in Figure~\ref{fig:Bicontour_C} \citepalias{Chamandy+Taylor15},
as well as the long tail extending out to $r\sim10\kpc$.
In the thicker disc model, Model~D, 
striking features are apparent at small radius, including a strong double-tail morphology in $\mbp$.
At $r=0$, $\mbr$ changes sign at about $z=\pm0.4\kpc$,
which is not very different from Model~A, in spite of the twice larger $h$ in Model~D.
However, real galaxies probably typically have $h$ equal to a few hundred $\pc$ at $r=0$, 
flaring to $\sim1\kpc$ for $r>10\kpc$ \citep{Chamandy+16},
which makes the small-$r$ behaviour of Model~D less relevant.
At larger radius $\gtrsim1\kpc$, the field resembles qualitatively that of Model~A,
though scaled to occupy the larger cross-section of the disc.
More interesting, perhaps,
is that $\mbr$ is less extended with radius than in Model~A,
while $\mbp$ is more extended.
Importantly, the overall field strength falls off less slowly with radius than for Model~A
because $\Dtilde\propto h^2$ remains above unity even at large $r$.

\subsection{Magnetic pitch angle}
\label{sec:pitch}
In panels~(g)--(i) of Figure~\ref{fig:Bcompare} and (c) of Figure~\ref{fig:Bcompare_D}, 
we plot the mean magnetic pitch angle 
averaged over the disc cross-section $\langle p \mean{B}^2\rangle/\langle \mean{B}^2\rangle$ 
in the saturated (solid) and kinematic (dashed) regimes.
The spikes at $r\sim1\kpc$ in Figure~\ref{fig:Bcompare}(i) and the peculiar behaviour of $p$ for $r\lesssim1\kpc$
are caused by $p$ wrapping around from $-90^\circ$ to $90^\circ$, as per the definition.
When averaging $p$ over the disc cross section, 
these discontinuities also lead to unphysical wiggles in $\langle p \mean{B}^2\rangle/\langle\mean{B}^2\rangle$ over the finite grid.
This is just a consequence of the discontinuous definition of $p$, not of any numerical problem in the simulation.
If $p$ were redefined to allow values $<-90^\circ$, 
$\langle p \mean{B}^2\rangle/\langle\mean{B}^2\rangle$ would continue to rise smoothly as $r\rightarrow0$.
Note also that the unphysical behaviour of $p\kin$ at $r\gtrsim7\kpc$ in panel~\ref{fig:Bcompare}(h) 
is not of any consequence as the field is effectively zero in that region.

As expected, $p<0$, $|p|$ is of the order of $10^\circ$ for the fiducial model, Model~A, 
and is slightly larger when an outflow is present, in Model~B.
Also as expected, 
$|p|$ in the saturated regime is generally smaller than that in the kinematic regime \citepalias{Chamandy+Taylor15}.

Blue dashed-dotted and dotted lines show the predictions of the analytical model 
for the saturated (equation~\eqref{psat}) and kinematic (equation~\eqref{pkin}) regimes, respectively.
Note that the analytical prediction for $p$ in the kinematic regime is independent of $V$,
whereas the numerical solution shows a weak dependence of $p$ on $V$ in the kinematic regime,
as can be seen by comparing the dashed lines in panels~(g) and (h) of Figure~\ref{fig:Bcompare}.
Analytical solutions predict $p$ remarkably accurately, especially for the saturated state.
This is not completely surprising as they were calibrated against local numerical saturated solutions \citepalias{Chamandy+Taylor15}.
However, $|p|$ at small $r$ in Model~C is larger than predicted because of the $\alpha^2$ effect.
Nevertheless, the agreement with global numerical solutions is striking 
and implies that the analytical solution and the approximations underlying it are quite robust.
Of course, solutions will in general depend on the galactic rotation curve,
and rotation curves that vary more rapidly in radius than the one adopted here
would lead to larger deviations of the global solutions from the local estimates,
as the latter ignore radial derivatives.

\subsection{Consequences of applying various approximations to the equations}
\label{sec:approx}
We study the effects of the various approximations using Figures~\ref{fig:Bicontour_A}--\ref{fig:Bicontour_D};
a concise summary of these approximations can be found in Table~\ref{tab:approximations}.
We have also studied plots of residuals between $\mbr$, $\mbp$, and $\mbz$ from any two different panels of a given figure, 
but for the sake of brevity, these are not shown.
Below, we sometimes refer to models run using different approximations as `sub-models';
in going from one panel (sub-model) to the next in sequence,
we make one additional approximation but otherwise keep the sub-model the same.
Panels~(a) show the full $\alpha^2\Omega$ solution. 
We first compare panels~(a) with panels~(b), which show the $\alpha\Omega$ solution.
This approximation neglects terms containing $\alpha$ in equation~\eqref{dTdt}.
For Model~A, shown in Figure~\ref{fig:Bicontour_A}, 
the $\alpha^2$ effect leads to a slight increase in the field components away from the midplane for $r\lesssim1\kpc$,
which is noticeable in the slight dip in $\mbp$ at $z=0$.
The same effect is seen in Model~B, but is less pronounced,
while for Model~C, the effect is visually similar, but here the $\alpha^2$ effect produces a decrement in $\mbp$ near the midplane
\citepalias[c.f.][]{Chamandy+Taylor15}, with an associated `push' of $\mbz$ to slightly larger radius.
For Model~D, the field strength at small $r$ is greatly reduced in panel~\ref{fig:Bicontour_D}(b)
compared to \ref{fig:Bicontour_D}(a),
and the effect on the morphology of $\mbp$ is qualitatively the same as for Model~A,
but more pronounced.

In going from panels~(b) to (c) of Figures~\ref{fig:Bicontour_A}--\ref{fig:Bicontour_D}, 
$\mbz$ is now neglected in the equations,
and then computed \textit{a posteriori} using $\bfDel\cdot\meanv{B}=0$.
This is known as the thin-disc approximation \citepalias{Ruzmaikin+88}, 
and is formally valid for $r\gg h$.
Terms involving $\del\mbz/\del z$ in the mean induction equation
are \textit{not} neglected in this approximation, however,
as $\del\mbz/\del z=-(1/r)\del(r\mbr)/\del r$ from $\bfDel\cdot\meanv{B}=0$.
In our implementation, this means neglecting radial derivatives of $\psi$,
\textit{except} in equation~\eqref{dpsidt}.
In fact, in going from panels~(b) to (c) equations~\eqref{dTdt} and \eqref{dpsidt}
are not affected (we confirm that kinematic solutions for $\mbr$ and $\mbp$ 
are almost identical), 
so any difference between the saturated solutions of these panels
stems from the non-linearity~\eqref{dalpha_mdt_formalism}.
For Models~A--C, there is a slight weakening of the field 
at small radius in going from panels~(b) to (c).
We note that the magnitude of $\mbr$, in particular, 
is significantly reduced at $r\lesssim2\kpc$, 
though its spatial distribution is only weakly affected.
Still, the thin-disc approximation can be seen to be rather robust for these models.
However, for the thick disc model, Model~D, departures are more significant,
as expected.

In going from panels~(c) to (d) of Figures~\ref{fig:Bicontour_A}--\ref{fig:Bicontour_D}, 
radial derivatives are now neglected: the local or slab approximation.
This leads to rather drastic cutoffs in $\mbr$ and $\mbp$ at small and large $r$,
where the local solutions are subcritical.
These cutoffs are of course unphysical.
The sharp cutoffs in $\mbr$ also lead to a poor reconstruction of $\mbz$.
It is also worth noting that $\mbp$ peaks at a slightly larger value when the slab approximation is made.
Nevertheless, these plots show that the slab approximation is quite valid for $r\gtrsim1\kpc$,
for the parameter values considered.
Perhaps fortuitously, 
slab solutions of panels~(d) better reproduce the strength of $\mbr$, near its peak,
than do thin-disc solutions of panels~(c).
The stronger $\mbr$ in panels~(d) as compared to panels~(c) is mainly due to the neglect
of the radial diffusion term in equation~\eqref{dpsidt} in the slab approximation:
re-running submodels~(c) with this term omitted leads to stronger $\mbr$ 
comparable with that of submodels~(b) and (d).

We next consider the effect of adopting the generalized algebraic quenching non-linearity
in place of the dynamical quenching formalism.
That is we use equation~\eqref{algebraic} with equation~\eqref{a_quench},
where $a$ evaluates to $2.5$, $1.8$, $0.6$, and $10.1$
for Models~A, B, C, and D, respectively.
Thus we compare panels~(d), which, like panels~(a)--(c), use dynamical quenching,
with panels~(e), which show solutions for a sub-model identical to that shown in panels~(d),
except now using generalized algebraic quenching.
The effect on the magnitude and morphology of $\meanv{B}$ is seen to be rather small.
For Model~B we see that the distributions of $\mbr$ and $\mbp$ broaden with respect to the midplane.
Further, $\mbr$ ceases to changes sign at the disc surface,
in agreement with the findings of \citetalias{Chamandy+14b}.

\subsection{Comparison of analytical solutions with numerical solutions}
Finally, we compare the above solutions with the analytical solutions of panels~(f) 
of Figures~\ref{fig:Bicontour_A}--\ref{fig:Bicontour_D},
calculated using the same resolution as that of the numerical grid.
That solution was obtained by normalizing the critical perturbation solution
using the critical no-$z$ solution, 
as explained in Section~\ref{sec:combined} and Table~\ref{tab:approximations}.
We have already seen from Figure~\ref{fig:Bcompare} that analytical no-$z$ solutions
reproduce rather faithfully the vertically averaged properties of the full solutions.
How well do 2.5D `no-$z$ + perturbation' solutions approximate full numerical solutions?
It can be seen that for all four models, 
the analytical solution of panels~(f) of Figures~\ref{fig:Bicontour_A}--\ref{fig:Bicontour_D}
agrees reasonably well with the full numerical solution of panels~(a),
but fail at small radius.
The differences between analytical and numerical solutions are more noticeable for Model~B than for Models~A and C, 
because of the limitations of the analytical methods for the case where large vertical outflows are present.
In the presence of a strong outflow, 
analytical solutions underestimate the radial extent of the magnetic field.
For the case with vanishing outflow, and presumably for weak outflows,
the analytical solution is as accurate as the numerical slab solution, shown in panels~(d), 
which neglects radial derivatives.
We note that standard estimates of the outflow velocity give $V$ of the order of $\sim10^{-2}$--$10^{-1}$ \citep{Shukurov+06}, 
so in this context the outflow in our Model~B of $V=0.25$ is very strong.
For Model~D, the thicker disk causes departures from the numerical solutions to be larger
than for Model~A, as expected.

\section{Discussion}
\label{sec:discussion}
The technique we have used to obtain a 2.5D analytical solution is simple (even seemingly naive).
Yet it gives rather accurate results, so long as we are not concerned with the very centre of the disc
and so long as outflows are not very strong. 
Why should this be?

\subsection{Local vs.~global}
\label{sec:local}
More specifically, how could it be that radially local axisymmetric solutions,
obtained at each radius by neglecting radial derivatives, 
can effectively be pasted together to form a global solution?
The field saturates first at the radius approximately corresponding to the peak 
of the eigenfunction in the linear regime,
and then proceeds to spread out. 
This spreading does not happen mainly by radial diffusion,
but by the much faster local growth of the field.
Therefore, saturation at each radius takes place almost independently \citep{Chamandy+13a}.

The exception to this picture is what takes place near the boundary of the dynamo-active region,
i.e.~at the radii $r_1$ and $r_2$, where $D=D\crit$.
There the local growth rate becomes negative and non-local radial processes become crucial
for reversing the balance, and rendering the dynamo supercritical.
Unfortunately, 
there does not seem to be a simple way to model the field saturation for $r<r_1$ and $r>r_2$,
so the field cuts off steeply (and unphysically) at $r_1$ and $r_2$.
Any improvement to the analytical model to allow for a more gradual tapering off of the field,
in better agreement with numerical solutions of panels~(a)--(c) 
of Figures~\ref{fig:Bicontour_A}--\ref{fig:Bicontour_D}, would be usefull.

\subsection{How well do the various approximations work?}
More broadly, our results explicitly demonstrate the viability 
of various approximations often used in the galactic dynamo literature,
while exposing some of their limitations.
The $\alpha\Omega$ approximation is shown to be reasonable for typical parameters,
but leads to an underestimate of $|p|$ at small radius, for instance.
For the rotationally saturated $\alpha\kin$ case of Model~C, 
and thicker disc of Model~D,
the $\alpha\Omega$ solution is inaccurate at small radius, where $\mbz$ is important.
Making, in addition, the thin disc approximation, 
for which $\mbz$ terms are neglected in the simulation,
and calculated \textit{post factum}, 
does not lead to a large difference in solutions 
for Models~A--C, which have a thin disc,
but $\mbr$ is significantly reduced in magnitude.
For Model~D, which has a thicker disc, 
such departures are even more significant, as would be expected.
The slab approximation pastes together local solutions that depend only parametrically on $r$.
As explained above, this solution is invalid near the radii where the local growth rate
transitions from positive to negative, but is otherwise remarkably accurate.
Replacing the dynamical quenching non-linearity by a generalized form of the algebraic $\alpha$-quenching non-linearity,
which is relatively easy to implement and less demanding computationally, 
is seen to have very little effect on the slab solution.
Finally, the analytical solution, which combines no-$z$ and perturbation solutions,
is surprisingly accurate, though not without limitations, 
as discussed in Sections~\ref{sec:results} and \ref{sec:local}.

\subsection{The disc centre}
As soon as one resorts to pasting together local solutions, 
the solution for the very central region of the disk becomes invalid.
This is far from ideal, but not particularly damaging, for at least three reasons.
First, the other approximations, such as the $\alpha\Omega$ approximation,
also lead to inaccuracies in this region.
Even the vacuum boundary conditions imposed are rather unjustified there,
as discussed in Section~\ref{sec:boundary}.
Further, other effects which have been neglected, such as the presence of the galactic bulge,
or a small bar in SAB galaxies, for example, would also be more important near the centre.
In summary then, we are probably losing less information
than it seems in the analytical model by `losing' the very centre.
Second, for potential applications, 
such as estimating the Faraday rotation due to intervening galaxies on distant AGN,
the surface area would effectively enter the calculation.
As $\sim1/15$ of the radius corresponds to only $\sim1/200$ of the surface area,
the effect of the very centre is likely to be negligible, 
given other potential uncertainties in the problem.
Third, ionized galactic discs are likely to be flared,
with $h$ typically somewhat smaller than $0.5\kpc$ at the centre.
This would allow the validity of all of the sub-models presented to extend down to a smaller radius $r\sim h$.

For the same reason, the analytical model also underestimates the field at larger radius $r\gtrsim r_2$.
The field in this region is relatively weak, though the relative surface area is significant.
The neglect of the field in this region is probably only a concern for models with strong outflows,
since the disparity is greater in that case,
as discussed in Section~\ref{sec:results}.

\subsection{Simple but powerful?}
We see then that the analytical solution, though crude, is reasonably accurate.
It reproduces faithfully certain key properties of the solution, 
such as the local vertically averaged pitch angle,
as well as the overall magnitude of each field component.
What makes this solution potentially powerful, 
however, compared with more accurate numerical solutions,
is the ease with which it can be implemented.
Such solutions could be painted onto all of the galaxies in a cosmological simulation 
almost instantaneously and almost effortlessly, to any desired spatial resolution.
The greatest shortcoming of such solutions may be
their neglect of non-local effects in the disc.
On the other hand, the various parameters can be made to depend on $r$ without complicating the method.
For example, making the disc flared so that $h$ depends on $r$ is a rather trivial extension.
It may also be worth pointing out that analytical solutions are far more transparent,
with regard to their interpretation, than their numerical counterparts.

\subsection{Limitations of a pure disc model}
The disc solutions presented can be seen as `basic' solutions for spiral galaxies, 
and they \textit{may} be rather sufficient 
to explain the regular magnetic fields of \textit{certain} galaxies.
However, many galaxies harbour strong, apparently large-scale, 
magnetic fields high above the ionized thin disc \citep{Krause14}.
To include this part of the galaxy in our models,
it is necessary to relax certain approximations, such as the $z$-independence of $l$, $u$, etc.,
that may be reasonably accurate for a disc, but fail drastically for a disc-halo system.
Such an extension to our models is a logical next step,
especially considering that haloes, being relatively extended, 
may be important in
the Faraday rotation of background sources.
Important work using mean-field dynamo simulations 
to explore large-scale magnetic fields of disc-halo systems has been carried out,
and provides a useful starting point
\citep{Sokoloff+Shukurov90,Brandenburg+92,Brandenburg+93,Moss+Sokoloff08,Moss+10}.
At the same time, 
analytical dynamo solutions for the halo are at present unavailable and would be interesting to develop.
Ultimately, the aim would be to model analytically the disc-halo system,
and compare solutions with more realistic numerical solutions,
as done above for the pure disc case.
We see the present work as a necessary `stepping stone' on the way to this goal.

\section{Conclusions}
\label{sec:conclusions}
We have presented a simple, effectively 2.5D axisymmetric analytical solution 
for the large-scale magnetic field of a disc galaxy in the saturated state.
This solution was shown to agree with full numerical solutions to a reasonable level of fidelity,
with the agreement extending down to the magnitudes 
and spatial morphologies of each individual component of the field.

Where our analytical solution is notably less reliable than full numerical solutions is at small radius, 
$r\lesssim1\kpc$ or smaller.
It is also less reliable if vertical outflow speeds from the disc are larger than, 
or in the upper range of, standard estimates,
or if the disc is thicker than standard estimates in the inner few $\kpc$.

Our solution is simple, transparent, and takes virtually no time to implement.
It is also versatile because the underlying parameters,
such as the turbulent speed of energy carrying eddies, disc scale height, 
and correlation time of turbulence, 
can be varied and can even be allowed to depend on radius.

In addition to obtaining numerical solutions for our rather general set of equations,
we also obtained numerical solutions from successively more idealized versions of these equations,
all of which are used to some extent in the dynamo literature.
By comparing such solutions we were able to assess the suitability of these approximations,
at least for the parameter values considered.

Possibly the greatest shortcoming of both the numerical and analytical models is the neglect of the gaseous halo,
which has been observed in several edge-on galaxies to emit polarized synchrotron radiation,
and probably contains large-scale magnetic field. 
Nevertheless, our model is, we believe, a solid base for exploring more realistic solutions in the future.
In the meantime, it is hoped that our analytical solution provides a viable starting point
for researchers who are interested in painting magnetic fields onto galaxies in order to explore various observational signatures.

\section*{Acknowledgments}
The author thanks K.~Subramanian for many useful discussions
during the development of the numerical code.
The author is also grateful to A.~Shukurov for providing comments on an early draft,
and to the anonymous referee for feedback
that significantly improved the presentation.

\footnotesize{
\noindent
\bibliographystyle{mn2e}
\bibliography{refs}
}

\appendix

\section{Full equations}
\label{sec:full_equations}
The toroidal and poloidal parts of equation~\eqref{dBdt} and equation~\eqref{dalpha_mdt} can respectively be written as
\begin{equation}
  \begin{split}
    \label{dTdt_full}
    \frac{\del T}{\del t}=&
      -r\frac{\del}{\del r}\left(\frac{\mur T}{r}\right) -\frac{\del}{\del z}(\muz T)
      +q\Omega\frac{\del\psi}{\del z} +\Sz\frac{\del\psi}{\del r}\\&
      -\alpha\Lambda^-\psi -\frac{\del\alpha}{\del r}\frac{\del\psi}{\del r} -\frac{\del\alpha}{\del z}\frac{\del\psi}{\del z}\\&
      +\eta\Lambda^-T +\frac{\del\eta}{\del r}\frac{\del T}{\del r} +\frac{\del\eta}{\del z}\frac{\del T}{\del z},
  \end{split}
\end{equation}
\begin{equation}
  \label{dpsidt_full}
  \frac{\del\psi}{\del t}= -\mur\frac{\del\psi}{\del r} -\muz\frac{\del\psi}{\del z} +\alpha T +\eta\Lambda^-\psi,
\end{equation}
\begin{equation}
  \begin{split}
    \label{dalpha_mdt_formalism_full}
    \frac{\del\alpha\magn}{\del t}=&
      -\frac{2\eta}{l^2r^2B\eq^2}\Biggl\{\alpha\left[\left(\frac{\del\psi}{\del r}\right)^2 
                                               +T^2 +\left(\frac{\del\psi}{\del z}\right)^2\right]\Biggr.\\&
      \Biggl.-\eta\left[\frac{\del\psi}{\del r}\frac{\del T}{\del r} 
                 -T\Lambda^-\psi +\frac{\del\psi}{\del z}\frac{\del T}{\del z}\right]\Biggr\}\\&
      -\frac{1}{r}\frac{\del}{\del r}(r\mur\alpha\magn) -\frac{\del}{\del z}(\muz\alpha\magn)\\&
      +\kappa\Lambda^+\alpha\magn 
      +\frac{\del\kappa}{\del z}\frac{\del\alpha\magn}{\del z} +\frac{\del\kappa}{\del r}\frac{\del\alpha\magn}{\del r},
  \end{split}
\end{equation}
where $\Sz\equiv r\del\Omega/\del z$ and
$\Lambda^\pm\equiv \del^2/\del r^2 \pm(1/r)\del/\del r +\del^2/\del z^2$.
In the present work, we have adopted $\mur=\Sz=\del\eta/\del r=\del\eta/\del z=\del\kappa/\del r=\del\kappa/\del z=0$.
This allows us to write these equations in the form~\eqref{dTdt}, \eqref{dpsidt}, and \eqref{dalpha_mdt_formalism}.

\label{lastpage}
\end{document}